\documentclass[12pt, draftclsnofoot, onecolumn]{IEEEtran}
\usepackage{cite}
\usepackage{amsmath,amssymb,amsfonts}
\usepackage{algorithmic}
\usepackage{graphicx}
\usepackage{textcomp}
\usepackage{xcolor}
\usepackage{subfigure}
\usepackage{booktabs}
\usepackage{epstopdf}
\usepackage{footnote}
\usepackage{amsmath}
\usepackage{algorithmic}
\usepackage{multicol}
\usepackage[caption=false,font=footnotesize]{subfig}
\usepackage{mathtools}

\usepackage{cite}
\ifCLASSINFOpdf
\else
\fi

\begin{document}
	\title{Hybrid Spherical- and Planar-Wave Channel Modeling and Estimation for Terahertz Integrated UM-MIMO and IRS Systems}
	\author{\IEEEauthorblockN{Yuhang~Chen, Renwang~Li,~\IEEEmembership{Student~Member,~IEEE,} Chong~Han,~\IEEEmembership{Member,~IEEE,} Shu~Sun,~\IEEEmembership{Member,~IEEE,} and Meixia~Tao,~\IEEEmembership{Fellow,~IEEE}
	}
		\thanks{
			This paper was presented in part at IEEE ICC, May 2022~\cite{ref_IRS_ICC_22}.

			Y. Chen and C. Han are with Terahertz Wireless Communications (TWC) Laboratory, Shanghai Jiao Tong University, Shanghai 200240, China (e-mail: \{yuhang.chen, chong.han\}@sjtu.edu.cn).
			
			R, Li, S. Sun and M. Tao are with Department of Electronic
			Engineering, Shanghai Jiao Tong University, Shanghai 200240, China (e-mail: \{renwanglee, shusun, mxtao\}@sjtu.edu.cn).

			}}
	\maketitle
	\thispagestyle{empty}
	\begin{abstract} 

Integrated ultra-massive multiple-input multiple-output (UM-MIMO) and intelligent reflecting surface (IRS) systems are promising for 6G and beyond Terahertz (0.1-10 THz) communications, to effectively bypass the barriers of limited coverage and line-of-sight blockage. 
However, excessive dimensions of UM-MIMO and IRS enlarge the near-field region, while strong THz channel sparsity in far-field is detrimental to spatial multiplexing. Moreover, channel estimation (CE) requires recovering the large-scale channel from severely compressed observations due to limited RF-chains. 
To tackle these challenges, a hybrid spherical- and planar-wave channel model (HSPM) is developed for the cascaded channel of the integrated system. 
The spatial multiplexing gains under near-field and far-field regions are analyzed, which are found to be limited by the segmented channel with a lower rank.
Furthermore, a compressive sensing-based CE framework is developed, including a sparse channel representation method, a separate-side estimation (SSE) and a dictionary-shrinkage estimation (DSE) algorithms. 
Numerical results verify the effectiveness of the HSPM, the capacity of which is only $5\times10^{-4}$ bits/s/Hz deviated from that obtained by the ground-truth spherical-wave-model, with 256 elements. 
While the SSE achieves improved accuracy for CE than benchmark algorithms, the DSE is more attractive in noisy environments, with 0.8~dB lower normalized-mean-square-error than SSE.

\boldmath
	\columnsep 0.2in
	\end{abstract}
\columnsep 0.2in


\section{Introduction}

Owning abundant bandwidth of multi-GHz up to even Terahertz (THz), the THz spectrum ranging from 0.1 to 10~THz has attracted upsurging attention from academia and industry in recent years.
The THz wireless communications have the capability to support Terabit-per-second high data rates, which are envisioned as a pillar candidate for 6G wireless networks~\cite{ref_THz_demand_IF1,ref_THz_demand_Rappaport, ref_THz_demand_ZhiChen}. 
However, the THz wave suffers from large free-space attenuation, strong molecular absorption, and high non-line-of-sight (NLoS) propagation losses incurred from reflection, scattering, and diffraction.
Therefore, it is challenging to achieve robust wireless transmission in complex occlusion environments, especially when line-of-sight (LoS) is blocked~\cite{ref_IRS_6G}.
Moreover, power amplifiers with low efficiency at THz frequencies have constrained output power, which results in the low reception signal-to-noise ratio (SNR) thus constraining the communication distance~\cite{ref_Distance_problem}.

To overcome the distance limitation, the ultra-massive multiple-input multiple-output (UM-MIMO) systems are exploited in the THz band~\cite{ref_Realizing_UM_MIMO}. Thanks to the sub-millimeter wavelength, hundreds and even thousands of antennas can be deployed in the UM-MIMO, which provides high array gain to compensate for the propagation losses. 
Furthermore, as a key technology to enable intelligent propagation environments in 6G systems, the intelligent reflecting surface (IRS) has been advocated in the literature~\cite{ref_IRS_survey_Wu,ref_IRS_survey_DRM_2020,ref_lrw_HBF_RIS}. 
The IRS is equipped with a metamaterial surface of the integrated circuit, which can be programmed to enable passive beamforming with high energy efficiency~\cite{ref_IRS_survey_Wu}.
At lower frequencies, the IRS is majorly used to increase the achievable data rates.
By contrast, in the THz band, the IRS can effectively bypass the barrier of the LoS blockage problem, by precisely controlling the reflection of incident THz signals~\cite{ref_THz_demand_IF1,ref_RIS_emil}. 
To combine, an integrated UM-MIMO and IRS systems can simultaneously solve the distance limitation and LoS blockage problems for THz wireless communications.

Channel modeling, analysis, and channel estimation (CE) arise as three inter-related open challenges of the THz integrated UM-MIMO and IRS systems.
First, while most existing work on channel modeling in IRS assisted systems only considers the far-field propagation~\cite{ref_model_emil_Rayleighmodel}, 
the near-field region is expanded with an enlarged dimension of antenna arrays in UM-MIMO and IRS, relative to the sub-millimeter wavelength of the THz wave.
The consideration of near-field spherical-wave propagation is imperatively needed~\cite{ref_HSPM,ref_Dai_Near_field}. 
Second, each segmented channel in the integrated IRS and UM-MIMO systems can be in near-field and far-field, whose multiplexing capability concerning the cascaded channel remains unclear. 
Moreover, due to the large reflection, scattering, and diffraction losses, the THz channel is generally sparse and dominated by a LoS and only a few NLoS paths~\cite{ref_Multiray}. As a result, the THz multi-antenna channels suffer from limited multiplexing capability imposed by the number of multi-paths instead of the number of antennas as in the microwave band. Therefore, the spatial multiplexing capability needs to be assessed and possibly enhanced in the THz integrated UM-MIMO and IRS systems. 

Third, the hybrid UM-MIMO structures with low hardware cost are commonly deployed in the THz systems, which exploit a much smaller number of RF-chains than antennas~\cite{ref_hybrid_beamforming}. 
This hybrid architecture is helpful to reduce power and hardware costs, which however causes a research problem for CE. That is, with the enormous amount of antennas in the UM-MIMO and passive reflecting elements lacking the signal processing ability of the IRS, CE has to recover a high-dimensional channel relating to the antennas and passive elements, from severely compressed low dimensional signal on the RF-chains. 
Moreover, the consideration of spherical-wave propagation alters the structure of channel models, leading that traditional solutions based on planar-wave propagation become ineffective. New CE methods to address these problems are thus needed. 

\subsection{Related Works}

\subsubsection{Channel Modeling and Analysis}

In the literature, mainly two categories of MIMO channel models are considered, namely, the spherical-wave model (SWM) and the planar-wave model (PWM),  which are effective in addressing the near-field and far-field effects, respectively~\cite{ref_spherical_fronts, ref_SW_PW_Modeling}.
As an improvement to PWM and SWM, we proposed a hybrid spherical- and planar-wave channel model (HSPM) for THz UM-MIMO systems in~\cite{ref_HSPM} , which accounted for PWM within the subarray and SWM among subarrays.
Compared to the PWM and SWM, the HSPM is more effective by deploying a few channel parameters to achieve high accuracy in the near-field. 
In the IRS assisted communication systems, an alternative physically feasible Rayleigh fading model was proposed in~\cite{ref_model_emil_Rayleighmodel} under the far-field assumption. 
By taking both near-field effect and IRS into consideration, the authors in~\cite{ref_modeling_analysis_THz_IRS} considered the SWM for THz integrated IRS and UM-MIMO systems. 
However, the SWM suffers from high complexity with the massive number of elements in the UM-MIMO and IRS~\cite{ref_HSPM}. To date, an effective model addressing the near-field effect in UM-MIMO and IRS systems is still required.

In the IRS systems, the channel analysis mainly focuses on sum rate, power gain, spectral efficiency (SE), and energy efficiency (EE). 
In microwave systems, the authors in~\cite{ref_IRS_capacity_characterization} characterized the capacity limit by jointly optimizing the IRS reflection coefficients and the MIMO transmit covariance matrix. The distribution and the outage probability of the sum rate were derived in~\cite{ref_analysis_rate_outage}, by considering the SWM of the LoS and PWM of the NLoS. 
A closed-form expression of the power gain was derived in~\cite{ref_emil_power_scaling}, and the near-field and far-field behaviors were analyzed. 
At higher frequencies, the ergodic capacity under the Saleh-Valenzuela model was derived and optimized in~\cite{ref_rwl_RIS_capacity}, while the SE and EE are analyzed in~\cite{ref_modeling_analysis_THz_IRS}.
As a critical metric to assess the spatial-multiplexing capability of the channel, the channel rank analysis has been conducted in the THz UM-MIMO systems. 
To enhance the limited spatial multiplexing in the THz UM-MIMO systems, a widely-spaced multi-subarray (WSMS) structure with enlarged subarray spacing was proposed in~\cite{ref_WSMS}, where the channel rank can be improved by a factor equal to the number of subarrays.
However, the rank analysis in the THz integrated UM-MIMO and IRS systems are still lacking in the literature.

\subsubsection{Channel Estimation}
CE for IRS assisted MIMO systems has been explored in the literature~\cite{ref_IRS_CE_DnDL,ref_IRS_CE_active_DL,ref_IRS_CE_2timescale,ref_IRS_CE_matrix_cali,ref_IRS_CE_ANM,ref_IRS_CE_Analysis,ref_IRS_CE_CE,ref_trice,ref_OMP_IRS,ref_IRS_CE_CS_THz}, which can be categorized into two main categories, namely, estimation of the segmented channels from user equipment (UE) to IRS and IRS to base station (BS), and estimation of the UE-IRS-BS cascaded channel. 
On one hand, since the passive IRS lacks signal processing capability, it is hard to directly separate each channel segment. 
Thus, the segmented CE schemes often require special hardware design, e.g., inserting active IRS elements or using full-duplex equipment, both of which however increase the hardware cost~\cite{ref_IRS_CE_DnDL,ref_IRS_CE_active_DL,ref_IRS_CE_2timescale,ref_IRS_CE_matrix_cali}. 
In~\cite{ref_IRS_CE_DnDL} and~\cite{ref_IRS_CE_active_DL}, a few IRS elements were activated during the pilot reception. The deep-learning tool was then assisted for CE with considerable estimation accuracy.
By deploying a full-duplex operated BS, a two timescale CE method was proposed in~\cite{ref_IRS_CE_2timescale}. 
The segmented CE problem was formulated as a matrix factorization problem and solved in~\cite{ref_IRS_CE_matrix_cali}, which can be operated with purely passive IRS. However, this scheme does not address the near-field effect.

On the other hand, since most precoding designs are based on the knowledge of the cascaded channel, the estimation of which has been explored in most existing schemes~\cite{ref_IRS_CE_ANM,ref_IRS_CE_Analysis,ref_IRS_CE_CE,ref_trice,ref_OMP_IRS,ref_IRS_CE_CS_THz}. 
In~\cite{ref_IRS_CE_ANM}, a two-stage atomic norm minimization problem was formulated, by which the super-resolution channel parameter estimation was conducted to efficiently obtain the channel-state-information. 
Theoretical analysis of the required pilot overhead and a universal CE framework were proposed in~\cite{ref_IRS_CE_Analysis}, which are effective in guiding the design of training and CE. 
However, all of them are limited to be applicable with fully digital MIMO structures. 
By exploiting the channel sparsity in the mmWave and THz bands, compressive sensing (CS) based CE methods in hybrid MIMO systems were explored in~\cite{ref_IRS_CE_CE,ref_trice,ref_OMP_IRS,ref_IRS_CE_CS_THz}. 
These schemes deploy the spatial discrete Fourier transform (DFT) based on-grid codebook to sparsely represent the channel, which is beneficial in achieving reduced training overhead. 
On the downside, the near-field effect was not incorporated in the DFT codebook, which results in limited estimation accuracy of these schemes in the near-field region of the THz multi-antenna systems~\cite{ref_HSPM}.

\subsection{Contributions}
To fill the aforementioned research gap, in this work, we first model the cascaded channel and study the spatial multiplexing in THz integrated UM-MIMO and IRS systems, by considering both near-field and far-field effects. 
Based on that, we propose a CS-based CE framework. 
In particular, we develop a subarray-based on-grid codebook to sparsely represent the channel.
Then, a separate side estimation (SSE) and a spatial correlation inspired dictionary shrinkage estimation (DSE) algorithms are proposed to realize low-complexity CE.
In our prior and shorter version~\cite{ref_IRS_ICC_22}, we proposed the cascaded channel model and analyzed the spatial multiplexing of the integrated systems. 
In this work, we further derive the on-grid codebook and propose two low-complexity CE algorithms. Furthermore, we perform  substantially more extensive performance evaluation. The major contributions of this work are summarized as follows.

 \begin{itemize}
 \item We propose an HSPM for the cascaded channel in the THz integrated UM-MIMO and IRS systems, and analyze the spatial multiplexing gain of the cascaded channel.
 By addressing both near-field and far-field effects. The proposed channel model accounts for the PWM within the subarray and SWM among subarrays, which achieves better accuracy than the PWM and lower complexity than the SWM.
 Moreover, the spatial multiplexing gain of the cascaded channel is analyzed when the segmented channels satisfy the near-field and far-field conditions, respectively. We prove that the rank of the cascaded channel is constrained by the individual channel with a lower rank. Furthermore, we present that spatial multiplexing can be improved based on the widely-spaced architecture design.

 \item We develop a CS-based CE framework including the sparse channel representation and sparse recovery algorithms.
 First, we propose a subarray-based codebook to sparsely represent the HSPM.
 Since the HSPM takes the subarray as a unit, by which each block is the sub-channel for a specific subarray pair, the proposed codebook possesses much higher accuracy than the traditional DFT codebook. 
 Based on this, we propose low complexity DSE and SSE sparse recovery algorithms for the CE of the integrated system.
 The SSE algorithm separately estimates the positions of non-zero grids on each side of the channel. By contrast, the DSE algorithm further reduces the complexity of SSE by exploring the fact that the angles for different subarray pairs are close in the spatial domain.

 \item We carry out extensive simulations to demonstrate the effectiveness of the proposed HSPM and CE methods.
 The HSPM can accurately capture the propagation features of the THz integrated UM-MIMO and IRS system. Numerically, we demonstrate that the capacity based on the HSPM is fairly close to that obtained by the ground-truth SWM.
 Moreover, both SSE and DSE achieve higher accuracy compared to existing algorithms. While SSE in general owns the highest accuracy, the DSE is more attractive at lower SNR, e.g., below 0~dB. 

 \end{itemize}
 
 The remainder of this paper is organized as follows. The system and channel models are introduced in Sec.~\ref{sec_System_Overview}. Spatial multiplexing analysis is presented in Sec.~\ref{sec_Spatial_Multiplexing_Analysis}.
 The subarray-based codebook and the SSE, DSE CE algorithms are proposed in~\ref{sec_Channel_Estimation}.
Extensive performance evaluation and numerical analysis are conducted in Sec.~\ref{sec_Performance_Evaluation}. Finally, the paper is concluded in Sec.~\ref{sec_Conclusion}.

\textbf{Notation}: 
$a$ is a scalar.
$\mathbf{a}$ denotes a vector. 
$\mathbf{A}$ represents a matrix. 
$\mathbf{A}(m,n)$ stands for the element at the $m^{\rm th}$ row and $n^{\rm th }$ column in $\mathbf{A}$.
$\mathbf{A}(i,:)$ depicts the $i^{\rm th}$ row of $\mathbf{A}$. 
$\mathbf{A}(:,j)$ refers to the $j^{\rm th}$ column of $\mathbf{A}$. 
$\mathbf{p}(m:n)$ denotes the $m^{\rm th}$ to $n^{\rm th}$ elements of $\mathbf{p}$.
$\mathbf{p}(m)$ refers to the $m^{\rm th}$ element of $\mathbf{p}$.
$(\cdot)^{\mathrm{T}}$ defines the transpose. 
$(\cdot)^{\mathrm{H}}$ refers to the conjugate transpose.
$(\cdot)^{\dag}$ denotes the pseudo inverse.
$|\cdot|$ depicts the absolute value. 
$\Vert\cdot\Vert_0$ defines the $l_0$-norm.
$\Vert\cdot\Vert_2$ stands for the $l_2$-norm.
$\mathbb{C}^{M\times N}$ depicts the set of $M\times N$-dimensional complex-valued matrices. 
$\otimes$ refers to Kronecker product. 
$\circ$ denotes Khatri-Rao product. 
$\propto$ depicts proportional sign.

 \section{System Overview}
 \label{sec_System_Overview}
 
\subsection{System Model}
\begin{figure}[t]
	\centering
	{\includegraphics[width= 0.8\textwidth]{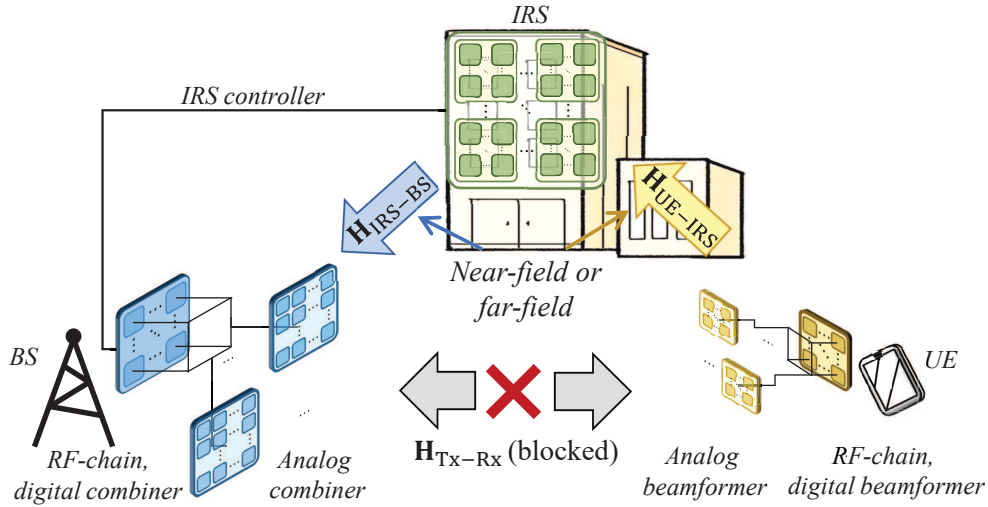}}
	\caption{Integrated THz UM-MIMO and IRS system.} 
	\label{fig_system_model}
	\vspace{-5mm}
\end{figure}

As illustrated in Fig.~\ref{fig_system_model}, we consider a THz integrated UM-MIMO and IRS communication system. 
The WSMS THz UM-MIMO with planar-shaped antenna arrays is equipped at both BS and UE. The direct channel between the BS and UE is considered to be blocked and inaccessible due to the occlusion propagation environment~\cite{ref_trice,ref_IRS_CE_ANM}. 
The communication link is assisted by a planar-shaped IRS with $M$ passive reflecting elements, which is connected to the BS via an IRS controller. Moreover, we consider that the IRS can be divided into $K_m$ planar-shaped subarrays, $M=K_mN_{am}$, where $N_{am}$ denotes the number of passive reflecting elements on each subarray.

In the WSMS design at the BS, $K_b$ subarrays are deployed, each of which contains $N_{ab}$ antennas. 
The total number of antennas is obtained as $N_{b} = K_bN_{ab}$.
On one hand, within the subarray, the antenna spacing $d = \lambda/2$, where $\lambda$ denotes the carrier wavelength. 
On the other hand, the subarray spacing is multiple times half-wavelengths~\cite{ref_WSMS}. 
Moreover, each subarray is connected to one RF-chain. 
In THz UM-MIMO systems, a much smaller number of RF-chains than the number of antennas is often adopted, for lower hardware cost and higher EE~\cite{ref_hybrid_beamforming}. Therefore, we have $ K_b\ll N_b$. 
Similarly, the UE is composed of $N_u$ antennas, which can be divided into $K_u$ subarrays, each of which is connected to one RF-chain. Each subarray contains $N_{au}$ antennas, satisfying $N_{u} = K_uN_{au}$ and $ K_u\ll N_u$.

By considering an uplink transmission, the received signal $\mathbf{y}\in\mathbb{C}^{N_{sb}}$ at the BS is denoted as
 \begin{equation}\label{equ_received_signal_IRS}
	\begin{split}
	\mathbf{y}&=\overline{\mathbf{W}}^{\mathrm{H}}
	\mathbf{H}^{\rm cas}\overline{\mathbf{F}}\mathbf{s}+\overline{\mathbf{W}}^{\mathrm{H}}\mathbf{n},
	\end{split}
\end{equation}
where $N_{sb}$ denotes the number of signalstreams st BS, $\overline{\mathbf{W}}=\mathbf{W}_{\rm RF}\mathbf{W}_{\rm BB}\in\mathbb{C}^{N_b\times N_{sb}}$ represents the combining matrix, with $\mathbf{W}_{\rm RF}\in\mathbb{C}^{N_b\times K_b}$ and $\mathbf{W}_{\rm BB}\in\mathbb{C}^{K_b\times N_{sb}}$ denoting the analog and digital combining matrices, respectively.
The cascaded channel matrix from the UE to BS is depicted as $\mathbf{H}^{\rm cas}\in\mathbb{C}^{N_b\times N_u}$. 
The beamforming matrix at the UE is represented as $\overline{\mathbf{F}}=\mathbf{F}_{\rm RF}\mathbf{F}_{\rm BB}\in\mathbb{C}^{N_u\times N_{su}}$, 
where $N_{su}$ depicts the transmitted number of signal streams at UE, $\mathbf{F}_{\rm RF}\in\mathbb{C}^{N_u\times K_u}$ and $\mathbf{F}_{\rm BB}\in\mathbb{C}^{K_u\times N_{su}}$ refer to the analog and digital beamforming matrices, respectively.
Moreover, $\mathbf{s}\in\mathbb{C}^{N_{su}}$ describes the transmitted signal, while $\mathbf{n}\in\mathbb{C}^{N_u}$ represents additive white Gaussian noise (AWGN). 
The analog beamforming and combining are completed by phase shifters. Therefore, each element of $\mathbf{W}_{\rm RF}$ and $\mathbf{F}_{\rm RF}$ satisfies constant module constraint, which can be expressed as $\mathbf{W}_{\rm RF}(i,j) = \frac{1}{\sqrt{N_b}}e^{jw_{i,j}}$, $\mathbf{F}_{\rm RF}(i,j) = \frac{1}{\sqrt{N_u}}e^{jf_{i,j}}$, where $w_{i,j},f_{i,j} \in[0,2\pi]$ denote the phase shift value. 
In addition, $\mathbf{W}_{\rm BB}$ and $\mathbf{F}_{\rm BB}$ are usually set as identity matrices during the training process for CE. In this case, there is $N_{su} = K_u$ and $N_{sb} = K_b$.

\subsection{Channel Model}
\label{subsec_Channel_Model}
The cascaded channel matrix $\mathbf{H}^{\rm cas}$ in~\eqref{equ_received_signal_IRS} can be represented as
\begin{equation}
	\label{equ_H_cas}
	\mathbf{H}^{\rm cas} = \mathbf{H}_{\rm IRS-BS}\overline{\mathbf{P}}\mathbf{H}_{\rm UE-IRS},
\end{equation}
where $\mathbf{H}_{\rm IRS-BS}\in\mathbb{C}^{N_b\times M}$ stands for the segmented channel from the IRS to BS,
$\overline{\mathbf{P}}={\rm{diag}}\{\mathbf{p}\}\in\mathbb{C}^{M\times M}$ denotes the passive beamforming matrix at the IRS, where $\mathbf{p} = [e^{j\tilde{p}_1},\dots,e^{j\tilde{p}_M}]^{\rm{T}}$, $\tilde{p}_m\in[0,2\pi]$ refers to the phase shift of the $m^{\rm th}$ element of the IRS, $m = 1,\dots, M$. 
In addition, $\mathbf{H}_{\rm UE-IRS}\in\mathbb{C}^{M\times N_u}$ depicts the segmented channel from the UE to IRS. 
The segmented channels can be characterized based on different modeling assumptions. The PWM and SWM are explored by addressing the far-field and near-field effects, respectively~\cite{ref_HSPM}. 
Particularly, the receiver (Rx) is in the far-field of the antenna array at the transmitter (Tx) when the communication distance $D$ is larger than the Rayleigh distance $\frac{2S^2}{\lambda}$, where $S$ denotes the array aperture. In this case, the wave is approximated to propagate in a plane and the PWM can be adopted. 
By contrast, the SWM has to be considered when the communication distance is smaller than the Rayleigh distance, where the Rx is located in the near-field and the propagation travels in a sphere. 
	
As an improvement to the PWM and SWM, we proposed the idea of HSPM in~\cite{ref_HSPM} in the THz UM-MIMO systems, which possesses less complexity than the SWM and achieves better modeling accuracy than the PWM in the near-field condition. In the following, we first introduce the PWM, SWM and HSPM for the segmented channels $\mathbf{H}_{\rm IRS-BS}$ and $\mathbf{H}_{\rm UE-IRS}$ in~\eqref{equ_H_cas}, respectively. Then, we propose the HSPM for the cascaded channel $\mathbf{H}^{\rm cas}$. 
To facilitate the description, during the introduction of different channel models, we use Tx to represent the IRS in $\mathbf{H}_{\rm IRS-BS}$ and the UE in $\mathbf{H}_{\rm UE-IRS}$, and use Rx to denote the BS in $\mathbf{H}_{\rm IRS-BS}$ and the IRS in $\mathbf{H}_{\rm UE-IRS}$, respectively.
Moreover, we consider that Tx is composed of $N_t$ elements and $K_t$ subarrays, while Rx employs $N_r$ antennas and $K_r$ subarrays, respectively.

\subsubsection{PWM}\label{subsec_PWM}
The PWM suitable for the far-field propagation region can be denoted as~\cite{ref_SW_PW_Modeling}
\begin{equation}\label{equ_PWM}
\mathbf{H}_{\rm P}=\Sigma_{p=1}^{N_p}\alpha_p
\mathbf{a}_{rp}\mathbf{a}_{tp}^{{\rm H}},
\end{equation}
where $\alpha_p$ represents the complex gain of the $p^{\rm th}$ propagation path, $p = 1,..., N_p$, with $N_p$ denoting the total number of paths. 
The array steering vectors at Rx and Tx are denoted as	$\mathbf{a}_{rp} = \mathbf{a}_{N_r}(\psi_{rpx},\psi_{rpz})\in\mathbb{C}^{N_r}$ and $\mathbf{a}_{tp} = \mathbf{a}_{N_t}(\psi_{tpx},\psi_{tpz})\in\mathbb{C}^{N_t}$, respectively.
Without loss of generality, by considering an $N$ element planar-shaped array on the x-z plane with physical angle pair $(\theta,\phi)$, the array steering vector $\mathbf{a}_N(\psi_{x},\psi_{z})\in\mathbb{C}^{N}$ can be expressed as
\begin{equation}\label{equ_array_steering_vector}
\mathbf{a}_N(\psi_{x},\psi_{z})=\left[1 \dots \mathrm{e}^{j\frac{2\pi }{\lambda}\psi_n}\dots \mathrm{e}^{j\frac{2\pi }{\lambda}\psi_N}\right]^{\mathrm{T}},
\end{equation}
where $\psi_n =\frac{d_{n_x}}{\lambda}\psi_x + \frac{d_{n_z}}{\lambda}\psi_z$, $\psi_x = {\rm sin}\theta{\rm cos}\phi $, $\psi_z = {\rm sin}\phi $ denotes the virtual angles, $d_{n_x}$ and $d_{n_z}$ stand for the distances between the $n^{\rm th}$ antenna to the first antenna on x- and z-axis, respectively. 
	
\subsubsection{SWM}\label{subsec_SWM}
The SWM is universally applicable to different communication distances, which individually calculates the channel responses of all antenna pairs between Tx and Rx to obtain the ground-truth channel. Due to the high complexity, the SWM is usually deployed in the near-field region, where the PWM becomes inaccurate. 
By denoting the communication distance of the $p^{\rm th}$ propagation path from the $n_t^{\rm th}$ transmitted antenna to the $n_r^{\rm th}$ received antenna as $D^{n_tn_r}_p$, the channel response of each antenna pair can be depicted as~\cite{ref_SW_PW_Modeling}
\begin{equation}\label{equ_SWM}
\mathbf{H}_{\rm S}(n_r,n_t)=\Sigma_{p=1}^{N_p}\lvert\alpha^{n_rn_t}_p\rvert e^{-j\frac{2\pi}{\lambda}D^{n_rn_t}_p},
 \end{equation} 
where $\mathbf{H}_{\rm S} \in\mathbb{C}^{N_r\times N_t}$ denotes the SWM channel matrix, $\alpha^{n_rn_t}_p$ represents the complex path gain. 

\subsubsection{HSPM}
The HSPM accounts for the PWM within one subarray and the SWM among subarrays, which can be denoted as~\cite{ref_HSPM}
\begin{equation}\label{equ_HSPM}
\begin{split}
	\mathbf{H}_{\rm HSPM}=\sum_{p=1}^{N_p}\vert\alpha_p\vert
	\left[\begin{array}{ccc}
		e^{-j\frac{2\pi}{\lambda}D^{11}_p}\mathbf{a}_{rp}^{11} (\mathbf{a}_{tp}^{11})^{\rm H}& \dots&e^{-j\frac{2\pi}{\lambda}D^{1K_t}_p}\mathbf{a}_{rp}^{1K_t}(\mathbf{a}_{tp}^{1K_t})^{\rm H}\\
	\vdots&\ddots&\vdots\\
		e^{-j\frac{2\pi}{\lambda}D^{K_r1}_p}\mathbf{a}_{rp}^{K_r1}(\mathbf{a}_{tp}^{K_r1})^{\rm H}& \cdots& e^{-j\frac{2\pi}{\lambda}D^{K_rK_t}_p}\mathbf{a}_{rp}^{K_rK_t}(\mathbf{a}_{tp}^{K_rK_t})^{\rm H}\\
	\end{array}\right],
\end{split}
	\end{equation}
where $D^{k_rk_t}_p$ stands for the distance between the $k_r^{\rm th}$ received and $k_t^{\rm th}$ transmitted subarray. The array steering vectors of the $p^{\rm th}$ path for the corresponding subarray pairs are denoted as $\mathbf{a}^{k_rk_t}_{rp} = \mathbf{a}_{N_{ar}}(\psi^{k_rk_t}_{rpx},\psi^{k_rk_t}_{rpz})$, and $\mathbf{a}_{tp}^{k_rk_t} = \mathbf{a}_{N_{at}}(\psi^{k_rk_t}_{tpx},\psi^{k_rk_t}_{tpz})$, respectively, which have similar forms as~\eqref{equ_array_steering_vector}.
The virtual angles $\psi^{k_rk_t}_{rpx} = {\rm sin}\theta^{k_rk_t}_{rp}{\rm cos}\phi^{k_rk_t}_{rp} $, $\psi^{k_rk_t}_{rpz} = {\rm sin}\phi^{k_rk_t}_{rp} $, $\psi^{k_rk_t}_{tpx} = {\rm sin}\theta^{k_rk_t}_{tp}{\rm cos}\phi^{k_rk_t}_{tp} $, $\psi^{k_rk_t}_{tpz} = {\rm sin}\phi^{k_rk_t}_{tp} $, where $(\theta^{k_rk_t}_{rp},\phi^{k_rk_t}_{rp})$ and $(\theta^{k_rk_t}_{tp},\phi^{k_rk_t}_{tp})$ stand for the azimuth and elevation angle pairs at Rx and Tx, respectively. Moreover, $N_{ar}$ and $N_{at}$ depict the number of antennas on the subarrays at Rx and Tx, respectively. 
We point out that the PWM and SWM are two special cases of HSPM when $K_t=K_r=1$ and $K_t=N_t$, $K_r=N_r$. 
In addition, the HSPM is accurate and can be adopted when the communication distance is smaller than the Rayleigh distance.

\subsection{HSPM for THz Integrated UM-MIMO and IRS Systems}

By replacing the segmented channels $\mathbf{H}_{\rm IRS-BS}$ and $\mathbf{H}_{\rm UE-IRS}$ of $\mathbf{H}^{\rm cas}$ in~\eqref{equ_H_cas} by the expression in~\eqref{equ_HSPM}, the HSPM for the cascaded channel $\mathbf{H}^{\rm cas}$ can be represented as
\begin{equation}
\label{equ_H_HSPM_IRS}
\begin{split}
	\mathbf{H}_{\rm HSPM}^{\rm cas} = \sum_{p_{i,b}=1}^{N_p^{\rm IB}}
	\vert\alpha_{p_{i,b}}\vert
	&\left[\begin{array}{ccc}
		\sum_{k_m=1}^{K_m} \mathbf{G}^{1k_m}_{p_{i,b}}\mathbf{E}^{k_m1}
		&\dots&
		\sum_{k_m=1}^{K_m}\mathbf{G}^{1k_m}_{p_{i,b}}\mathbf{E}^{k_mK_u}\\
		\vdots&\dots&\vdots\\
		\sum_{k_m=1}^{K_m} \mathbf{G}^{K_bk_m}_{p_{i,b}}\mathbf{E}^{k_m1}
		& \dots & \sum_{k_m=1}^{K_m} \mathbf{G}^{K_bk_m}_{p_{i,b}}\mathbf{E}^{k_mK_u}\\
		\end{array}\right],
\end{split}
\end{equation}
where $\alpha_{p_{i,b}}$ denotes the path gain for the $p_{i,b}^{\rm th}$ path of $\mathbf{H}^{\rm IRS-BS}$, $p_{i,b} = 1,\dots,N_p^{\rm IB}$, $N_p^{\rm IB}$ refers to the number of propagation paths in $\mathbf{H}^{\rm IRS-BS}$. 
The matrix $\mathbf{G}_{p_{i,b}}^{k_bk_m}\in\mathbb{C}^{N_{ab}\times N_{am}}$ is represented as
\begin{equation}\label{equ_G_pib}
	\mathbf{G}_{p_{i,b}}^{k_bk_m} = e^{-j\frac{2\pi}{\lambda}D^{k_bk_m}_{p_{i,b}}}\mathbf{a}_{rp_{i,b}}^{k_bk_m} (\mathbf{a}_{tp_{i,b}}^{k_bk_m})^{\rm H}\tilde{\mathbf{P}}^{k_m}, 
\end{equation}
where $D^{k_bk_m}_{p_{i,b}}$ stands for the communication distance between the $k_b^{\rm th}$ subarray at the BS and $k_m^{\rm th}$ subarray at the IRS for the $p_{i,b}^{\rm th}$ path.
Moreover, the received and transmitted array steering vectors are denoted as $\mathbf{a}_{rp_{i,b}}^{k_bk_m} = \mathbf{a}_{N_{ab}}(\psi_{rp_{i,b}x}^{k_bk_m},\psi_{rp_{i,b}z}^{k_bk_m})$ and $\mathbf{a}_{tp_{i,b}}^{k_bk_m} = \mathbf{a}_{N_{am}}(\psi_{tp_{i,b}x}^{k_bk_m},\psi_{tp_{i,b}z}^{k_bk_m})$ as~\eqref{equ_array_steering_vector}. The virtual angles $\psi_{rp_{i,b}x}^{k_bk_m}\! = \!{\rm sin} \theta_{rp_{i,b}}^{k_bk_m} {\rm cos} \phi_{rp_{i,b}}^{k_bk_m},$ $\psi_{rp_{i,b}z}^{k_bk_m}\! = \!{\rm sin}\phi_{rp_{i,b}}^{k_bk_m},$ $\psi_{tp_{i,b}x}^{k_bk_m}\! = \!{\rm sin} \theta_{tp_{i,b}}^{k_bk_m} {\rm cos} \phi_{tp_{i,b}}^{k_bk_m},$ $\psi_{tp_{i,b}z}^{k_bk_m} \!= \!{\rm sin}\phi_{tp_{i,b}}^{k_bk_m},$ $(\theta_{rp_{i,b}}^{k_bk_m},\phi_{rp_{i,b}}^{k_bk_m})$ and $(\theta_{tp_{i,b}}^{k_bk_m},\phi_{tp_{i,b}}^{k_bk_m})$ represent the physical angles pairs. The passive beamforming matrix of the $k_m^{\rm th}$ subarray at IRS is denoted as $\tilde{\mathbf{P}}^{k_m} = {\rm diag}\{\mathbf{p}(k_mN_{am}+1: (k_m+1)N_{am})\}$.

In~\eqref{equ_H_HSPM_IRS}, the matrix $\mathbf{E}^{k_mk_u}\in\mathbb{C}^{N_{am}\times N_{au}}$ can be expressed as
\begin{equation}\label{equ_E_kmku}
	\mathbf{E}^{k_mk_u} =
	\sum_{p_{u,i}}^{N_p^{\rm UI}}
	\vert\alpha_{p_{u,i} } \vert e^{-j\frac{2\pi}{\lambda}D^{k_mk_u}_{p_{u,i}}}\mathbf{a}_{rp_{u,i}}^{k_mk_u} (\mathbf{a}_{tp_{u,i}}^{k_mk_u})^{\rm H}, 
\end{equation}
where $N_p^{\rm UI}$ denotes the number of propagation paths in $\mathbf{H}_{\rm UE-IRS}$, $p_{u,i} = 1,\dots,N_p^{\rm UI}$, $\alpha_{p_{u,i} }$ represents
the path gain for the $p_{u,i}^{\rm th}$ path. Moreover, $D^{k_mk_u}_{p_{u,i}}$ stands for the communication distance between the $k_m^{\rm th}$ subarray at IRS and $k_u^{\rm th}$ subarray at UE.
The array steering vectors owning similar forms as~\eqref{equ_array_steering_vector} are denoted as
$\mathbf{a}_{rp_{u,i}}^{k_mk_u}=\mathbf{a}_{N_{am}}(\psi_{rp_{u,i}x}^{k_mk_u},\psi_{rp_{u,i}z}^{k_mk_u})$ and $\mathbf{a}_{tp_{u,i}}^{k_mk_u}=\mathbf{a}_{N_{au}}(\psi_{tp_{u,i}x}^{k_mk_u},\psi_{tp_{u,i}z}^{k_mk_u})$, 
where $\psi_{rp_{u,i}x} = {\rm sin} \theta_{rp_{u,i}}^{k_mk_u} {\rm cos} \phi_{rp_{u,i}}^{k_mk_u} $, $\psi_{rp_{u,i}z} = {\rm sin} \phi_{rp_{u,i}}^{k_mk_u} $, $\psi_{tp_{u,i}x} = {\rm sin} \theta_{tp_{u,i}}^{k_mk_u} {\rm cos} \phi_{tp_{u,i}}^{k_mk_u} $, $\psi_{tp_{u,i}z} = {\rm sin} \phi_{tp_{u,i}}^{k_mk_u} $, $(\theta_{rp_{u,i}}^{k_mk_u},\phi_{rp_{u,i}}^{k_mk_u})$ and $(\theta_{tp_{u,i}}^{k_mk_u},\phi_{tp_{u,i}}^{k_mk_u})$ stand for the angle pairs at IRS and BS, respectively.

Based on~\eqref{equ_G_pib} and~\eqref{equ_E_kmku}, the $(n_{ab},n_{au})^{\rm th}$ element for the production of $\mathbf{G}^{k_bk_m}_{p_{i,b}}\mathbf{E}^{k_mk_u}\in\mathbb{C}^{N_{ab}\times N_{au}}$ in~\eqref{equ_H_HSPM_IRS} can be represented as
\begin{equation}
	\label{equ_DE_detial}
	\begin{split}
	&(\mathbf{G}^{k_bk_m}_{p_{i,b}}\mathbf{E}^{k_mk_u})(n_{ab},n_{au}) = \sum_{p_{u,i}}^{N_p^{\rm UI}} \vert\alpha_{p_{u,i} } \vert
	e^{-j\frac{2\pi}{\lambda}(D^{k_bk_m}_{p_{i,b}} + D^{k_mk_u}_{p_{u,i}})}\times
	\\
	&\sum_{n_{am}=k_m N_{am} + 1}^{(k_m+1) N_{am}}
	\exp\left[ -j\pi\left(
	\zeta_{rp_{i,b}n_{ab}}^{k_bk_m} -
	\zeta_{tp_{i,b}n_{am}}^{k_bk_m} +
	\zeta_{tp_{u,i}n_{au}}^{k_mk_u} -
	\zeta_{rp_{u,i}n_{am}}^{k_mk_u} +
	e^{j\tilde{p}_{n_{am}}} 
	\right)\right],
	\end{split}
\end{equation}
where the aggregated phase $\zeta_{rp_{i,b}n_{ab}}^{k_bk_m}$ can be denoted as 
\begin{equation}
\zeta_{rp_{i,b}n_{ab}}^{k_bk_m} = (n_{abx} -1) \psi_{rp_{i,b}x}^{k_bk_m}+(n_{abz}-1)\psi_{rp_{i,b}z}^{k_bk_m},
\end{equation}
$n_{ab} = n_{abx}n_{abz}= 1,\dots, N_{ab}$, with $ n_{abx}$ and $ n_{abz}$ index the positions of the element at the subarray of UE on x- and z-axis, respectively.
Similarly, the aggregated phases $\zeta_{tp_{i,b}n_{am}}^{k_bk_m},$ 
$\zeta_{tp_{u,i}n_{ab}}^{k_mk_u}$ and 
$\zeta_{rp_{u,i}n_{am}}^{k_mk_u}$ in~\eqref{equ_H_HSPM_IRS} can be expressed as
\begin{subequations}
	\begin{align}
		\zeta_{tp_{i,b}n_{am}}^{k_bk_m}&= (n_{amx} - 1)\psi_{tp_{i,b}z}^{k_bk_m} +(n_{amz}-1){\rm sin}\psi_{rp_{i,b}z}^{k_bk_m},
		\\
		\zeta_{tp_{u,i}n_{ab}}^{k_mk_u}&=(n_{aux}-1)\psi_{tp_{u,i}x}^{k_mk_u}+(n_{auz}-1)\psi_{tp_{u,i}z}^{k_mk_u},
		\\
		\zeta_{rp_{u,i}n_{am}}^{k_mk_u}&=(n_{amx}-1)\psi_{rp_{u,i}x}^{k_mk_u}+(n_{amz}-1)\psi_{rp_{u,i}z}^{k_mk_u},
	\end{align}
\end{subequations}
where $n_{am} = n_{amx}n_{amz}= 1,\dots, N_{am}$, $n_{au} = n_{aux}n_{auz}= 1,\dots, N_{ab}$. 

 \section{Spatial Multiplexing Gains Analysis}
 \label{sec_Spatial_Multiplexing_Analysis}
 The cascaded channel $\mathbf{H}^{\rm cas}$ in~\eqref{equ_H_cas} is composed of two channel segments, $\mathbf{H}_{\rm IRS-BS}$ and $\mathbf{H}_{\rm UE-IRS}$. 
 Under near-field and far-field conditions, the segmented channels can adopt different channel models. 
 In this section, we analyze the spatial multiplexing capability of $\mathbf{H}^{\rm cas}$ in terms of channel rank, under different cases of the segmented channels. 

\subsection{Ranks of PWM, SWM, and HSPM}
To facilitate the analysis, we consider an end-to-end channel during the process of analyzing the ranks of PWM, SWM and HSPM and use Tx and Rx to represent each end. 
The number of propagation paths between Tx and Rx is denoted as $N_p$.
As studied in~\cite{ref_SW_PW_Modeling}, ranks of the PWM in~\eqref{equ_PWM} and SWM in~\eqref{equ_SWM} equal to $N_p$ and $N=\min\{N_t,N_r\}$, respectively.
To illustrate the rank of the HSPM in~\eqref{equ_HSPM}, we present \textit{\textbf{Lemma~1}}. 
\subsubsection*{\textbf{Lemma 1}} The rank of $\mathbf{H}_{\rm HSPM}$ in~\eqref{equ_HSPM} satisfies $\min\{K_rN_p, K_tN_p,N_r, N_t\}
\leq {\rm Rank}(\mathbf{H}_{\rm HSPM})\leq\min\{K_rK_tN_p,N_r, N_t \}$. 

\textit{Proof:} 
The dimension of the channel matrix $\mathbf{H}_{\rm HSPM}$ in~\eqref{equ_HSPM} is $N_r\times N_t$, the maximum rank of the channel is $\min\{N_r,N_t\}$. 
To prove the right-hand side inequality, we first consider for a fixed transmit subarray $k_t$ and propagation path $p$, elements in the set of array steering vectors $\left\{\mathbf{a}_{tp}^{1k_t}, \dots, \mathbf{a}_{tp}^{K_rk_t}\right\}$
are linearly independent.
Due to the distinguishability of propagation paths, the angles of different paths are different.
Therefore, by fixing transmit and receive subarrays as $k_r$ and $k_t$, respectively, $\mathbf{a}_{tp}^{k_rk_t}$ for different propagation paths $p = 1,\dots,N_p$ are linearly independent.
 
 We consider that the $n_r^{\rm th}$ received antenna is the $n_{ar}^{\rm th}$ element on the $k_r^{\rm th}$ received subarray.
 Therefore, the $n_r^{\rm th}$ row of the HSPM channel in~\eqref{equ_HSPM} $\mathbf{H}_{\rm HSPM}(n_r,:)$ can be expressed as
 \begin{equation}
 \label{equ_H_HSPMline}
 \mathbf{H}_{\rm HSPM}(n_r,:)=
	 \bigg[
	 \sum_{p=1}^{N_p}\beta_p^{k_r1}\left[ \mathbf{a}_{rp}^{k_r1}(n_{ar}) (\mathbf{a}_{tp}^{k_r1})^{\rm H}\right],\dots,
	 \sum_{p=1}^{N_p}\beta_p^{k_rK_t}\left[ \mathbf{a}_{rp}^{k_rK_t}(n_{ar}) (\mathbf{a}_{tp}^{k_rK_t})^{\rm H}\right]
		\bigg].
\end{equation}
Each row of $\mathbf{H}_{\rm HSPM}$ is a linear combination of $K_rK_tN_p$ linearly independent vectors as
 \begin{subequations}
	\label{equ_independent_vectors}
 \begin{align}
 &\Big[(\mathbf{a}_{tp}^{k_r1})^{\rm H},\mathbf{0},\dots,\mathbf{0}\Big],\\
 &\Big[\mathbf{0},(\mathbf{a}_{tp}^{k_r2})^{\rm H},\mathbf{0},\dots,\mathbf{0}\Big],\\
 &~~~~~~~~~~\dots\\
 &\Big[\mathbf{0},\dots,\mathbf{0},(\mathbf{a}_{tp}^{k_rK_t})^{\rm H}\Big],
 \end{align}
 \end{subequations}
 where $ k_r = 1,\dots,K_r$ and $\mathbf{0}$ is an all-zero vector of dimension $1\times N_{at}$.
 
 However, the angles of different paths to different received subarrays might be the same, leading that vectors in~\eqref{equ_independent_vectors} can be linearly dependent, which reduces the rank of the HSPM channel. Thus, there is ${\rm Rank}(\mathbf{H}_{\rm HSPM})\leq \min\{K_rK_tN_p,N_r, N_t \}$. To prove the left-hand side inequality, we consider an extreme case. For a fixed propagation path, the angles among different subarray pairs between Tx and Rx are same. In this case, the HSPM equals to the channel model in~\cite{ref_WSMS}, whose rank has been proved to be equal to $\min\{K_rN_p, K_tN_p,N_r, N_t \}$, which lower bounds the rank of the HSPM. 
Till here, we have completed the proof for \textit{\textbf{Lemma~1}}. $\hfill\blacksquare$

\subsection{Cascaded Channel Rank Analysis}
To analyze the rank of the cascaded channel, we first introduce the following lemma.
\subsubsection*{\textbf{Lemma 2}} For matrices $\mathbf{A}\in\mathbb{C}^{M\times N}$, $\mathbf{B}\in\mathbb{C}^{N\times N}$ and $\mathbf{C}\in\mathbb{C}^{N\times K}$, where $\mathbf{B}$ is a diagonal matrix, and ${\rm rank} (\mathbf{A}) = R_a$, ${\rm rank}(\mathbf{B}) = N$, ${\rm rank} (\mathbf{C}) = R_c$, we have
\begin{equation} \label{equ_rank_ineq}
{\rm rank}(\mathbf{ABC}) \leq \min\{R_a, R_c\}, 
\end{equation}
where the equality holds when $\mathbf{A}$ and $\mathbf{C}$ are full-rank matrices. 
 
 \textit{Proof:} 
 Since $\mathbf{B}$ is full row rank, we have ${\rm rank} (\mathbf{AB}) = {\rm rank}(\mathbf{A}) = R_a$. Then, ${\rm rank} (\mathbf{ABC}) \leq \min\{{\rm rank}(\mathbf{AB}), {\rm rank}(\mathbf{C})\} = \min\{R_a, R_c\}$. 
 When $\mathbf{A}$ is a full-rank matrix, we have ${\rm rank} (\mathbf{AB}) = {\rm rank}(\mathbf{A}) = N$. When $\mathbf{C}$ is a full-rank matrix, we have ${\rm rank} (\mathbf{ABC}) = {\rm rank} (\mathbf{AB})=N $. $\hfill\blacksquare$

Next, we analyze the rank of the cascaded channel. 
We adopt the PWM in the far-field region, while the SWM and HSPM are deployed for the near-field region, respectively.
 \subsubsection{Both segmented channels satisfy the far-field condition}
 In this case, both $\mathbf{H}_{\rm IRS-BS}$ and $\mathbf{H}_{\rm UE-IRS}$ in~\eqref{equ_H_cas} adopt PWM, whose ranks equal to $N_{p}^{\rm IB}$ and $N_{p}^{\rm UI}$, respectively.
 By denoting $\mathbf{H}^{\rm cas}$ in~\eqref{equ_H_cas} as $\mathbf{H}^{\rm cas}_{\rm PWM}$, 
 from \textit{\textbf{Lemma~2}}, we can state that ${\rm rank} (\mathbf{H}^{\rm cas}_{\rm PWM}) \leq \min\{N_{p}^{\rm IU},N_{p}^{\rm BI}\}$. Moreover, when $N_{p}^{\rm IU}=N_{p}^{\rm BI}=N_p$, ${\rm rank} (\mathbf{H}^{\rm cas}_{\rm PWM}) \leq N_{p}$. 

\subsubsection{One of the segmented channels satisfies the far-field condition, while the other satisfies the near-field condition} 
We first consider $\mathbf{H}_{\rm IRS-BS}$ satisfies the far-field condition and adopts the PWM, while $\mathbf{H}_{\rm UE-IRS}$ meets the near-field condition, which deploys SWM or HSPM. 
Since the number of propagation paths in the THz channel is much smaller than the number of elements in the UM-MIMO and IRS, ${\rm rank} (\mathbf{H}_{\rm IRS-BS}) =N^{\rm IB}_p < {\rm rank} (\mathbf{H}_{\rm UE-IRS})$. From \textit{\textbf{Lemma~2}}, we can obtain that ${\rm rank} (\mathbf{H}^{\rm cas}) \leq N_{p}^{\rm IB}$. 
A similar deduction can be drawn when $\mathbf{H}_{\rm IRS-BS}$ meets the near-field condition while $\mathbf{H}_{\rm UE-IRS}$ satisfies the far-field condition. 
Thus, when $N_{p}^{\rm IU}=N_{p}^{\rm BI}=N_p$, there is ${\rm rank} (\mathbf{H}^{\rm cas}) \leq N_{p}$. 

\subsubsection{Both segmented channels satisfy the near-field condition}

We denote $\mathbf{H}^{\rm cas}$ as $\mathbf{H}^{\rm cas}_{\rm SWM}$ when both $\mathbf{H}_{\rm IRS-BS}$ and $\mathbf{H}_{\rm UE-IRS}$ meet the near-field condition and adopt SWM. In this case, $\mathbf{H}_{\rm IRS-BS}$, $\overline{\mathbf{P}}$ and $\mathbf{H}_{\rm UE-IRS}$ are full-rank matrices. 
Based on \textit{\textbf{Lemma~2}}, we know that
${\rm rank} (\mathbf{H}^{\rm cas}_{\rm SWM}) = \min\{M,N_u,N_b\}$. When $N_u = N_b =M = N$, ${\rm rank} (\mathbf{H}^{\rm cas}_{\rm SWM}) = N$. Similarly, we denote $\mathbf{H}^{\rm cas}$ as $\mathbf{H}^{\rm cas}_{\rm HSPM}$ when both $\mathbf{H}_{\rm IRS-BS}$ and $\mathbf{H}_{\rm UE-IRS}$ adopt HSPM. By considering that $K_m=K_u=K_b=K$ and $N_p^{\rm IU}=N_p^{\rm BI}=N_p$, 
we have $ {\rm rank} (\mathbf{H}^{\rm cas}_{\rm HSPM}) \leq K^2N_p$.

From the above analysis, we can state that in the THz integrated UM-MIMO and IRS systems, the total rank of the cascaded channel is limited by the segmented channel with a smaller rank.
This suggests that the rank of the cascaded channel is increased only when both segmented channels meet the near-field condition, which inspires us to enlarge the array size and obtain a larger near-field region. It is worth noticing that the above discussions are not dependent on the IRS beamforming matrix $\overline{\mathbf{P}}$. 
Therefore, we further claim that given fixed segmented channels, the channel rank can not be improved by the IRS.

We will show in Sec.~\ref{sec_Performance_Evaluation} that the capacity of the THz integrated UM-MIMO and IRS system based on HSPM is close to that based on the ground truth SWM, which reveals the accuracy of the HSPM. 
In addition, the HSPM possesses lower complexity compared to the SWM~\cite{ref_HSPM}. Therefore, we directly adopt the HSPM for both segmented channels during the CE process.

\section{Channel Estimation} 
\label{sec_Channel_Estimation}
In this section, we present the CS-based CE framework for the THz integrated UM-MIMO and IRS communication systems, which is composed of three steps, namely, \textit{on-grid sparse channel representation}, \textit{signal observation} and \textit{sparse recovery algorithm}.
Specifically, the sparse channel representation is based on an on-grid codebook, by which the channel matrix is expressed as the production of the codebook and a sparse matrix. 
We first introduce the traditional DFT codebook, which is shown to be ineffective in the considered integrated systems.
Inspired by this, we propose a subarray-based codebook by considering the characteristic of the HSPM channel, which possesses higher sparsity and accuracy than the DFT codebook. 
Second, we introduce the training procedure to obtain the channel observation and formulate the CE problem as a sparse recovery problem. 
Third, to obtain the CE result, we develop the low-complexity SSE algorithm with high accuracy. The spatial correlation inspired DSE algorithm is further developed, which possesses lower complexity compared to the SSE, at the cost of slightly degraded accuracy.

\subsection{On-grid Sparse Channel Representation}

\subsubsection{Traditional DFT-based Sparse Channel Representation} 

In the literature, the spatial DFT-based on-grid codebook is widely deployed~\cite{ref_IRS_CE_CE,ref_trice,ref_OMP_IRS,ref_IRS_CE_CS_THz}. 
This codebook treats the entire antenna array as a unit, and considers that the virtual spatial angles $\psi_x = {\rm sin}\theta {\rm cos}\phi $ and $\psi_z = {\rm sin}\phi $ are taken form a uniform grid composed of $N_x$ and $N_z$ points, respectively. $\theta$ and $\phi$ denote the azimuth and elevation angles, while $N_x$ and $N_z$ refer to the number of antennas on x- and z-axis, respectively. 
In this way, the channel is sparsely represented as 
\begin{equation}\label{equ_on_grid_DFT}
	\mathbf{H} = \mathbf{A}_{\rm Dr} \boldsymbol{\Lambda}_{\rm D} \mathbf{A}_{\rm Dt}^{\rm H}
\end{equation}
where $\mathbf{A}_{\rm Dr} \in\mathbb{C}^{N_r \times N_r}$ and $\mathbf{A}_{\rm Dz} \in\mathbb{C}^{N_t \times N_t}$ refer to the two-dimensional DFT on-grid codebooks at Rx and Tx, respectively, which hold a similar form, and can be represented as $\mathbf{A}_{\rm D} =\Big[\mathbf{a}_{N}(-1, -1)\dots \mathbf{a}_{N}\left( \frac{2(n_x-1)}{N_x}-1, \frac{2(n_z-1)}{N_z}-1 \right) $ $\dots \mathbf{a}_{N_x}\left( \frac{2(N_x-1)}{N_x}-1, \frac{2(N_z-1)}{N_z}-1 \right)\Big]$. The sparse on-grid channel with complex gains on the quantized spatial angles is depicted by $\boldsymbol{\Lambda}_{\rm D} \in\mathbb{C}^{N_r\times N_t}$. 

\begin{figure}[t]
	\setlength{\belowcaptionskip}{0pt}
	\centering
	 \subfigure[Amplitude of the on-grid channel in compact array using the DFT codebook.]{\includegraphics[width=0.3\textwidth]{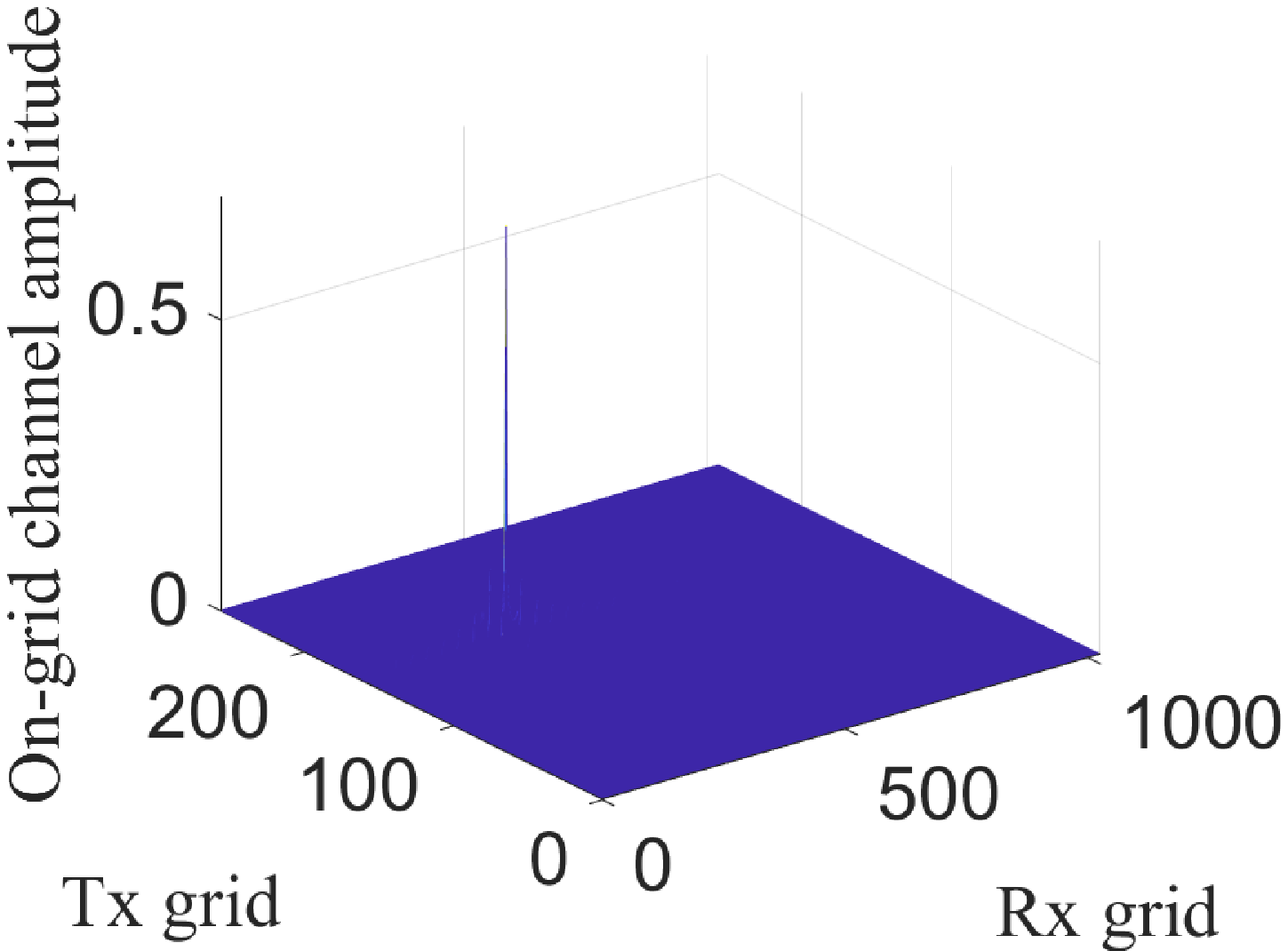} }
		 \subfigure[Amplitude of the on-grid channel in the WSMS using the DFT codebook, the subarray spacing is $64\lambda$.]{
		\includegraphics[width=0.3\textwidth]{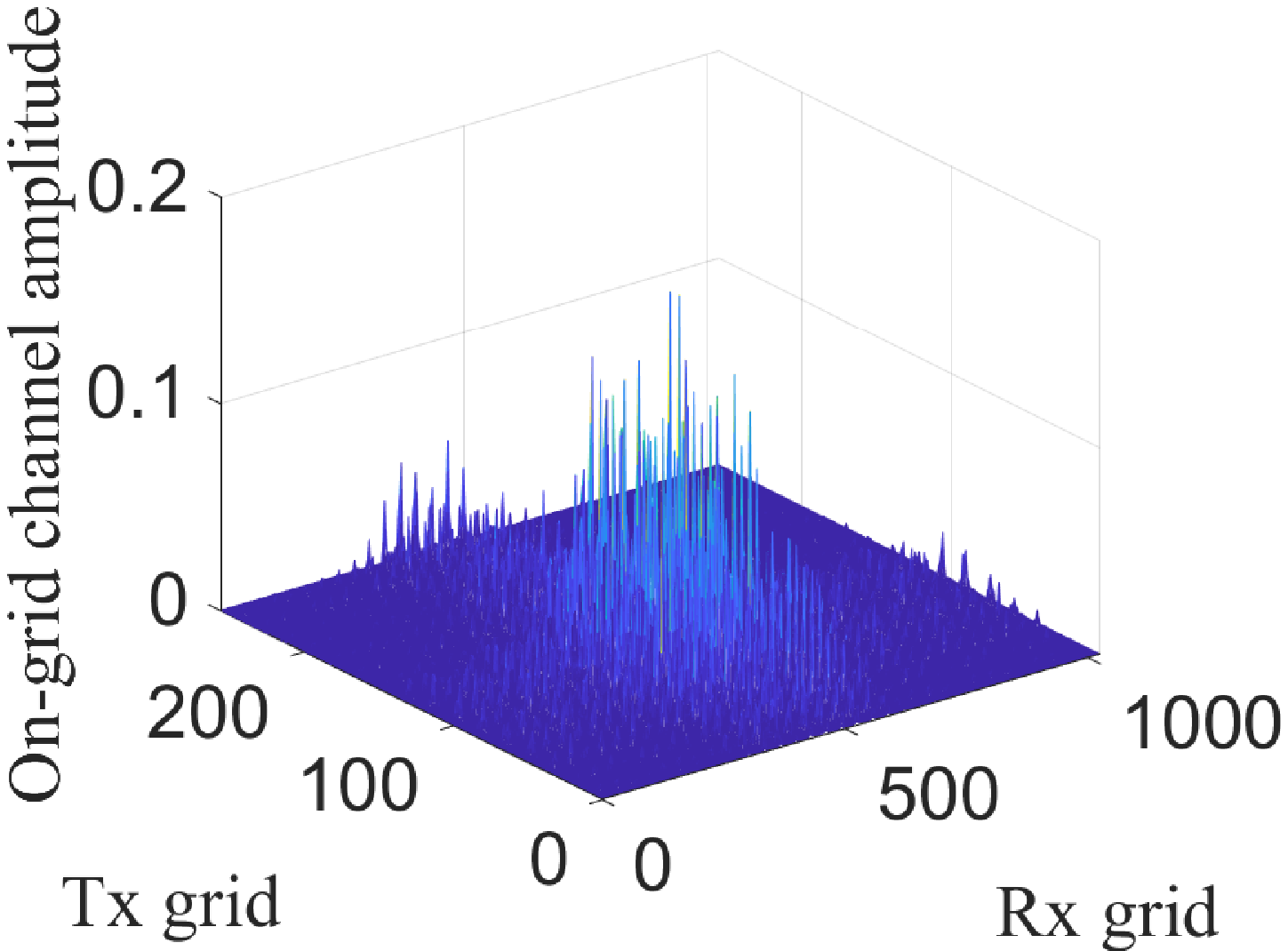} }
		\subfigure[Amplitude of the on-grid channel in the WSMS using the subarray-based codebook, the subarray spacing is $64\lambda$.]{
		\includegraphics[width=0.3\textwidth]{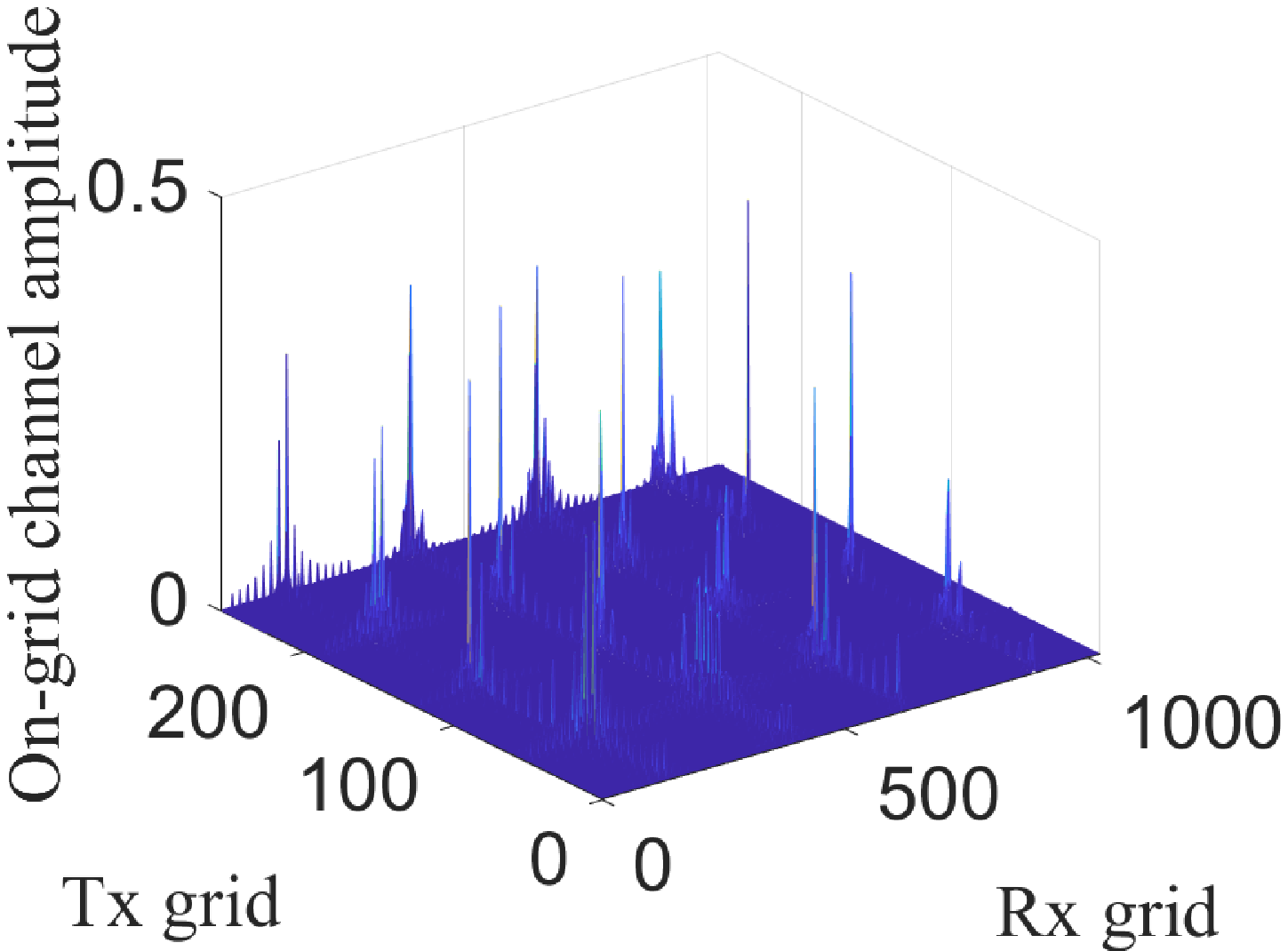} }
		\caption{Amplitude of on-grid channels in the compact array and WSMS with different codebooks by considering HSPM, $N_t = 256$, $N_r = 1024$, $K_t=K_r=4$. The number of propagation path is 1. }
		\label{fig_ongrid_channels}
		\vspace{ -5mm}
\end{figure}

To assess the performance of the DFT codebook, we first evaluate the sparsity of on-grid channel $\boldsymbol{\Lambda}_{\rm D}$ in~\eqref{equ_on_grid_DFT} in different cases, by considering the HSPM. 
Moreover, since in practice, there does not exist a grid whose amplitude is strictly equal to 0,
we consider that the sparsity of the on-grid channel equals the number of grids whose amplitude is greater than a small value, e.g., 0.01. 
First, as illustrated in Fig.~\ref{fig_ongrid_channels}(a), the amplitude of $\boldsymbol{\Lambda}_{\rm D}$ is shown by considering a compact array without enlarging the subarray spacing. The on-grid channel is sparse, the number of grids with an amplitude larger than 0.01 is only 397, which is much smaller than the preset total number of grids, i.e., 262144. 
By contrast, in Fig.~\ref{fig_ongrid_channels}(b), the amplitude of the on-grid channel in the WSMS is plotted. 
The on-grid channel contains 2755 grids with amplitude larger than 0.01. Therefore, the DFT codebook lacks sparsity in representing the HSPM. 

\subsubsection{Proposed Subarray-based Sparse Channel Representation} 
We observe that the HSPM in~\eqref{equ_HSPM} views each subarray as a unit, each block of which is the production of the array steering vectors for the subarrays at Rx and Tx, respectively. Inspired by this, we consider a subarray-based on-grid codebook. 
At Rx, the virtual spatial angles for each subarray are considered to be taken from fixed $N_{ar} = N_{arx}N_{arz}$ grids, where $N_{arx}$ and $N_{arz}$ refer to the number of elements on x- and z-axis of the subarray at Rx, respectively. 
The corresponding DFT codebook is expressed as 
$\mathbf{U}_{\rm Dr} = 
\Big[\mathbf{a}_{N_{ar}}(-1, -1),$ $\dots\!  ,\mathbf{a}_{N_{ar}}\left( \frac{2(n_{arx}-1)}{N_{arx}}-1, \frac{2(n_{arz}\!-\!1)}{N_{arz}}-1 \!\right)\!,
$
$\dots ,\mathbf{a}_{N_{ar}}\left( \frac{2(N_{arx}-1)}{N_{arx}}-1, \frac{2(N_{arz}-1)}{N_{arz}}-1 \right)\Big],$ $n_{arx} = 1,\dots,N_{arx}$, $n_{arz} = 1,\dots,N_{arz}$. 
We define $\overline{\mathbf{A}}_{\rm r} \in \mathbb{C}^{N_r\times N_r}$ as the subarray-based codebook at Rx, which deploys $K_r$ $\mathbf{U}_{\rm Dr}$ on its diagonal as
\begin{equation} \label{equ_subacodebook_rx}
	\overline{\mathbf{A}}_{\rm r} = {\rm blkdiag}\left[ \mathbf{U}_{\rm Dr},\dots, \mathbf{U}_{\rm Dr}\right]. 
\end{equation}
The on-grid codebook matrix at Tx $\overline{\mathbf{A}}_{\rm t} \in\mathbb{C}^{N_t \times N_t}$ is constructed similarly. 
Therefore, the on-grid representation of the HSPM in~\eqref{equ_HSPM} based on the subarray-based codebook can be denoted as 
\begin{equation}\label{equ_HSPM_grid}
    \mathbf{H}_{\rm HSPM} \approx \overline{\mathbf{A}}_{\rm r}\overline{\boldsymbol{\Lambda}}
    \overline{\mathbf{A}}_{\rm t}^{\rm H},
\end{equation}
where $\overline{\boldsymbol{\Lambda}}\in\mathbb{C}^{N_r \times N_t} $ is a sparse matrix.
If all spatial angles were taken from the grids and not equal to each other, $\overline{\boldsymbol{\Lambda}}$ would contain $K_rK_tN_p$ non-zero elements.

The amplitude of the on-grid channel $\overline{\boldsymbol{\Lambda}}$ in~\eqref{equ_HSPM_grid} using the proposed codebook is plotted in Fig.~\ref{fig_ongrid_channels}(c), by considering the same channel as in Fig.~\ref{fig_ongrid_channels}(b).
The number of grids with an amplitude larger than 0.01 is 1609, which is 1164 smaller than that by using the DFT codebook in Fig.~\ref{fig_ongrid_channels}(b). 
In addition, to reveal the accuracy of the on-grid channel, we calculate the difference between the real channel ${\mathbf{H}}_{\rm HSPM}$ and the reconstructed channels approximated by the on-grid channel and the codebooks in~\eqref{equ_on_grid_DFT} and~\eqref{equ_HSPM_grid} as ${\frac{  \left\Vert  \mathbf{A}_{\rm Dr} \boldsymbol{\Lambda}_{\rm D} \mathbf{A}_{\rm Dt}^{\rm H} - {\mathbf{H}}_{\rm HSPM}\right\Vert_2^2  } { { \left\Vert {\mathbf{H}}_{\rm HSPM}\right\Vert_2^2}  }}$ and
${\frac{  \left\Vert \overline{\mathbf{A}}_{\rm r}\overline{\boldsymbol{\Lambda}}
\overline{\mathbf{A}}_{\rm t}^{\rm H} - {\mathbf{H}}_{\rm HSPM}\right\Vert_2^2  } { { \left\Vert {\mathbf{H}}_{\rm HSPM}\right\Vert_2^2}  }}
$, respectively. 
The approximation error based on the proposed codebook is around 4 dB lower than that based on the DFT codebook. 
To this end, we state that the proposed codebook is more efficient than the traditional DFT codebook, which possesses higher sparsity and lower approximation error.

\subsection{Training Process and Problem Formulation}
\label{subsec_training_process}

\begin{figure}[t]
	\centering
	{\includegraphics[width= 0.9\textwidth]{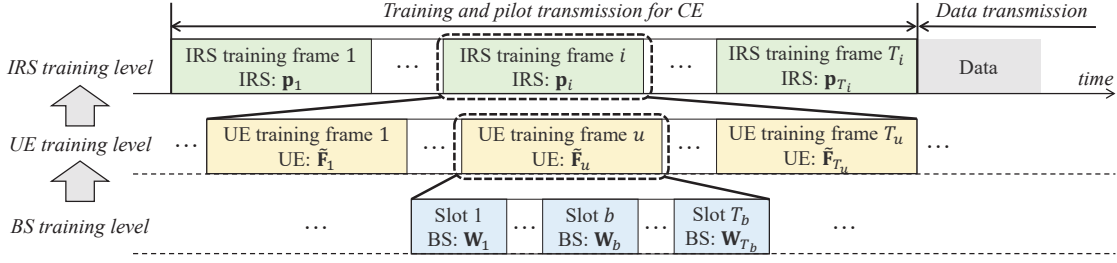}}
	\caption{Illustration of the training process.} 
	\label{fig_training_framework}
	\vspace{-5mm}
\end{figure}

\subsubsection{Training Process for Channel Observation}
We consider an uplink pilot training procedure, as illustrated in Fig.~\ref{fig_training_framework}, which is conducted in three levels, namely, the BS training, UE training and IRS training, respectively. 
During the training process, the UE transmits the known pilot signals to the BS via the IRS in $T = T_bT_uT_i$ training slots for uplink CE, where $T_b$, $T_i$ and $T_u$ denote the number of training slots for the BS, IRS and UE, respectively. 
At the $(b,u,i)^{\rm th}$ slot, the UE deploys the training beamformer $\tilde{\mathbf{F}}_{u} \in\mathbb{C}^{N_u\times N_{su}}$ and transmits the pilot signal $\mathbf{s}_{b,u,i}\in\mathbb{C}^{N_{su}}$, $b=1,\dots, T_b$, $u=1,\dots, T_u$, $i=1,\dots, T_i$. 
In the meantime, the training phase shift vector $\mathbf{p}_{i} \in\mathbb{C}^{M}$ and combiner $\mathbf{W}_{b} \in\mathbb{C}^{N_b \times N_{sb}}$ are deployed at the IRS and BS, respectively, to obtain the received signal ${\mathbf{y}}_{b,u,i}\in\mathbb{C}^{N_{sb}}$ at the BS, which can be represented as
\begin{equation}
 \label{equ_received_signal_biu}
 {\mathbf{y}}_{b,u,i} = \mathbf{W}_{b}^{\rm H} \mathbf{H}_{\rm IRS-BS }{\rm diag} \{\mathbf{p}_{i}\} \mathbf{H}_{\rm UE-IRS}\tilde{\mathbf{F}}_{u} \mathbf{s}_{b,u,i} + {\mathbf{n}}_{b,u,i},
\end{equation}
where ${\mathbf{n}}_{b,u,i} = \mathbf{W}_{b}^{\rm H}\tilde{\mathbf{n}}_{b,u,i}\in\mathbb{C}^{N_{sb} }$, and $\tilde{\mathbf{n}}_{b,u,i}\in\mathbb{C}^{N_{b}}$ refers to the received AWGN. 

The BS training is first conducted, in which totally $T_b$ different training combiners are used to obtain the received signal as~\eqref{equ_received_signal_biu}. By collecting $\mathbf{y}_{b,u,i}, b = 1,\dots, T_b$ as $\mathbf{y}_{u,i} = [\mathbf{y}_{1,u,i}^{\rm T},
\dots,\mathbf{y}_{T_b,u,i}^{\rm T} ]^{\rm T} \in\mathbb{C}^{N_{sb} T_b }$, the received signal after BS training can be expressed as 
\begin{equation}
	\label{equ_received_signal_BS_training}
		\mathbf{y}_{u,i} 
		=\left(\mathbf{f}_u^{\rm T} \otimes \mathbf{W}^{\rm H}\right) \mathbf{H}^{\rm mul} \mathbf{p}_{i} + \mathbf{n}_{u,i},
\end{equation}
where $\mathbf{f}_{u} = \tilde{\mathbf{F}}_{u} \mathbf{s}_{b,u,i} \in \mathbb{C}^{N_u}$ stands for the equivalent training beamformer, $\mathbf{W} = [\mathbf{W}_{1},\dots, \mathbf{W}_{T_b}]\in\mathbb{C}^{N_b \times N_{sb} T_b}$ denotes the training combiner. The multiplied channel matrix $\mathbf{H}^{\rm mul} = \mathbf{H}_{\rm UE-IRS}^{\rm T} \circ \mathbf{H}_{\rm IRS-BS} \in\mathbb{C}^{N_uN_b\times M}$, and 
$ \mathbf{n}_{u,i} =[\mathbf{n}_{1,u,i}^{\rm T},\dots, \mathbf{n}_{T_b,u,i}^{\rm T}]^{\rm T} \in\mathbb{C}^{ N_{sb} T_b}$ represents the collected noise.

After one round of BS training, the UE changes its beamformer $\tilde{\mathbf{F}}_{u}$ to complete the UE training. Particularly, totally $T_u$ beamformers are used to obtain the received signal as~\eqref{equ_received_signal_BS_training}. By collecting $\mathbf{y}_{u,i}$ for $u=1,\dots, T_u$ as $\mathbf{y}_{i} = [\mathbf{y}_{1,i}^{\rm T}, \dots,\mathbf{y}_{T_u,i}^{\rm T} ]^{\rm T} \in\mathbb{C}^{N_{sb} T_b T_u }$, we can obtain
\begin{equation}
	\label{equ_received_signal_UE_training}
		\mathbf{y}_{i} 
		=\left(\mathbf{F}^{\rm T} \otimes \mathbf{W}^{\rm H}\right) \mathbf{H}^{\rm mul} \mathbf{p}_{i} + \mathbf{n}_{i},
\end{equation}
where $\mathbf{F} = [\mathbf{f}_{1},\dots, \mathbf{f}_{T_u}]\in\mathbb{C}^{N_u\times T_u}$ denotes the UE training beamforming matrix. Moreover, $ \mathbf{n}_{i} =[\mathbf{n}_{1,i}^{\rm T},\dots, \mathbf{n}_{T_u,i}^{\rm T}]^{\rm T} \in\mathbb{C}^{ N_{sb} T_b T_u}$ represents the noise. 
Finally, the phase shift vector of the IRS $\mathbf{p}_{i}$ is changed to conduct the IRS training. 
After obtaining each $\mathbf{y}_{i}$ as~\eqref{equ_received_signal_UE_training}, $i=1,\dots,T_i$, we stack $\mathbf{y}_{i}$ as $\mathbf{Y} = [\mathbf{y}_{1},\dots, \mathbf{y}_{T_i}]\in\mathbb{C}^{N_{sb}T_bT_u\times T_i}$, which can be represented as 
\begin{equation}
 \label{equ_received_signal_IRS_training}
 \mathbf{Y} = \left( \mathbf{F}^{\rm T} \otimes \mathbf{W}^{\rm H} \right) \mathbf{H}^{\rm mul} \mathbf{P} + \mathbf{N},
\end{equation}
where $ \mathbf{P} = [\mathbf{p}_{1},\dots, \mathbf{p}_{T_i}] \in \mathbb{C}^{M\times T_i}$ refers to the training phase shift matrix. In addition, $\mathbf{N} = [\mathbf{n}_{1},\dots, \mathbf{n}_{T_i}]\in\mathbb{C}^{N_{sb}T_bT_u\times T_i}$ represents the stacked noise. 

In this work, CE refers to estimating the multiplied channel matrix $\mathbf{H}^{\rm mul}$ in~\eqref{equ_received_signal_IRS_training}. 
Based on the proposed codebook in~\eqref{equ_HSPM_grid}, $\mathbf{H}^{\rm mul}$ in~\eqref{equ_received_signal_IRS_training} can be represented as 
\begin{subequations}
	\label{equ_Hmul_ongrid_initial}
	\begin{align}
			\mathbf{H}^{\rm mul}&\approx\left( \overline{\mathbf{A}}_{\rm tUE-IRS}^{*}\overline{\boldsymbol{\Lambda}}_{\rm UE-IRS}^{\rm T}\overline{\mathbf{A}}_{\rm rUE-IRS}^{\rm T}\right) \circ 
			\left( \overline{\mathbf{A}}_{\rm rIRS-BS}\overline{\boldsymbol{\Lambda}}_{\rm IRS-BS}\overline{\mathbf{A}}_{\rm tIRS-BS}^{\rm H} \right),\\
			&=\mathbf{A}_{\rm r} \tilde{\boldsymbol{\Lambda}} \tilde{\mathbf{A}}_{\rm t},
	\end{align}
\end{subequations}
where $\overline{\mathbf{A}}_{\rm tUE-IRS} \in\mathbb{C}^{N_u\times N_u}$ and $\overline{\mathbf{A}}_{\rm rUE-IRS} \in\mathbb{C}^{M\times M} $ denote the codebook matrices for the UE-IRS channel at UE and IRS, respectively, $\overline{\boldsymbol{\Lambda}}_{\rm UE-IRS} \in\mathbb{C}^{M\times N_u}$ denotes the sparse on-grid channel.
The codebook matrices for the IRS-BS channel at the BS and IRS are denoted as $\overline{\mathbf{A}}_{\rm rIRS-BS} \in\mathbb{C}^{N_b\times N_b}$ and $\overline{\mathbf{A}}_{\rm tIRS-BS} \in\mathbb{C}^{M\times M}$, respectively. $\overline{\boldsymbol{\Lambda}}_{\rm IRS-BS} \in\mathbb{C}^{N_b\times M} $ stands for the corresponding sparse matrix. 
Moreover, $\mathbf{A}_{\rm r} = \left( \overline{\mathbf{A}}_{\rm tUE-IRS}^{*} \otimes \overline{\mathbf{A}}_{\rm rIRS-BS} \right) \in\mathbb{C}^{N_uN_b\times N_uN_b} $ stands for the combined codebook matrix at the left-hand side, $\tilde{\boldsymbol{\Lambda}} = \left( \overline{\boldsymbol{\Lambda}}_{\rm UE-IRS}^{\rm T} \otimes \overline{\boldsymbol{\Lambda}}_{\rm IRS-BS} \right) \in\mathbb{C}^{N_uN_b\times M^2} $ depicts the multiplied sparse matrix, $\tilde{\mathbf{A}}_{\rm t} = \left(\overline{\mathbf{A}}_{\rm rUE-IRS}^{\rm T} \circ \overline{\mathbf{A}}_{\rm tIRS-BS}^{\rm H} \right) \in\mathbb{C}^{M^2\times M}$ represents the combined transmit codebook matrix.

It is worth noticing that $\overline{\mathbf{A}}_{\rm rUE-IRS}=\overline{\mathbf{A}}_{\rm tIRS-BS}^{\rm H} = \overline{\mathbf{A}}_{\rm IRS}$, where $\overline{\mathbf{A}}_{\rm IRS}$ denotes the on-grid codebook matrix at the IRS. Therefore, the multiplied channel matrix can be transformed as 
\begin{equation}
\label{equ_Hmul_ongrid}
		\mathbf{H}^{\rm mul}\approx \mathbf{A}_{\rm r} {\boldsymbol{\Lambda}} {\mathbf{A}}_{\rm t},
\end{equation}
where ${\boldsymbol{\Lambda}} \in\mathbb{C}^{N_uN_b\times M}$ denotes the sparse on-grid channel mstrix, which is a function of $\mathbf{A}_{\rm r}$, $\tilde{\mathbf{A}}_{\rm t}$ and $\tilde{\boldsymbol{\Lambda}}$, ${\mathbf{A}}_{\rm t} = {\mathbf{A}}_{\rm IRS} \in\mathbb{C}^{M\times M}$ denotes the codebook matrix on the right-hand side. 
In addition, we point out that the rows of the non-zero elements in ${\boldsymbol{\Lambda}}$ corresponds to the grid points in ${\mathbf{A}}_{\rm r}$, while the columns of non-zero elements in ${\boldsymbol{\Lambda}}$ indicate the grid points in ${\mathbf{A}}_{\rm t}$.

\subsubsection{Problem Formulation}
By combining the on-grid channel representation in~\eqref{equ_Hmul_ongrid} with the channel observation in~\eqref{equ_received_signal_IRS_training}, we can obtain
\begin{equation}
\label{equ_received_signal_rewrite}
	\mathbf{Y} = \left( \mathbf{F}^{\rm T} \otimes \mathbf{W}^{\rm H} \right) \mathbf{A}_{\rm r} {\boldsymbol{\Lambda}} {\mathbf{A}}_{\rm t} \mathbf{P} + \mathbf{N}. 
\end{equation}
The CE problem can be formulated as a sparse signal recovery problem as
\begin{subequations}\label{equ_deisgn_problem_matrix}
	\begin{align}
		&{\min}~\left\Vert\boldsymbol{\Lambda}\right\Vert_0,\\
		&{\rm s.t.}~\left\Vert\mathbf{Y} -\left( \mathbf{F}^{\rm T} \otimes \mathbf{W}^{\rm H} \right) \mathbf{A}_{\rm r} {\boldsymbol{\Lambda}} {\mathbf{A}}_{\rm t} \mathbf{P}\right\Vert_0 \leq \epsilon, 
	 \end{align}
\end{subequations}
where $\epsilon$ is a constant to measure the estimation error. 
In addition, the $l_0$ norm in problem~\eqref{equ_deisgn_problem_matrix} is usually transformed into the $l_1$ norm, due to its non-convexity~\cite{ref_OMP}. 

To solve the problem in~\eqref{equ_deisgn_problem_matrix}, the received signal $\mathbf{Y}$ can be vectorized as $\mathbf{y}_{\rm vec} = {\rm vec}\{\mathbf{Y}\} \in\mathbb{C}^{N_{sb}T}$ to obtain
$\mathbf{y}_{\rm vec}=\tilde{\boldsymbol{\Phi}} \tilde{\boldsymbol{\Psi}} \mathbf{h} + \mathbf{n}_{\rm vec}$,
where $\tilde{\boldsymbol{\Phi}} = \left( \mathbf{P}^{\rm T} \otimes \mathbf{F}^{\rm T} \otimes \mathbf{W}^{\rm H} \right) \in\mathbb{C}^{N_{sb}T \times N_uN_bM}$ defines the measurement matrix, the overall codebook matrix is $\tilde{\boldsymbol{\Psi}} = \left( \mathbf{A}_{\rm t}^{\rm T}\otimes \mathbf{A}_{\rm r} \right)\in\mathbb{C}^{N_uN_bM \times N_uN_bM }$.
Moreover, $\mathbf{h} = {\rm vec}\{\boldsymbol{\Lambda} \} \in\mathbb{C}^{N_uN_bM} $ is a sparse vector containing the complex gains on the grids of the codebook, $\mathbf{n}_{\rm vec} = {\rm vec} \{\mathbf{N}\} \in\mathbb{C}^{N_{sb}T}$ represents the vectorized noise. Various of greedy algorithms such as orthogonal matching pursuit (OMP)~\cite{ref_OMP_IRS} and compressive sampling matching pursuit (CoSaMP)~\cite{ref_CoSaMP} can be used to recover $\mathbf{h} $ from $\mathbf{y}_{\rm vec}$. 
However, the dimension of $\tilde{\boldsymbol{\Psi}}$ is proportional to the number of antennas at BS $N_b$, UE $N_u$ and the number of passive reflecting elements at IRS $M$.
In our considered UM-MIMO and IRS systems, the dimension becomes unacceptably large and the computational complexity of the existing greedy algorithms upsurges.

\subsection{Sparse Recovery Algorithms}

\label{subsec_Sparse_recovery_algorithms}

\begin{table}[t]
	\centering
	\renewcommand
	\arraystretch{} 
	\begin{tabular}{l}
		\toprule
		\textbf{Algorithm 1:} SSE Algorithm \\
		\midrule 
		\textbf{Input}: Received signal $\mathbf{Y}$ in~\eqref{equ_received_signal_rewrite}, combined training matrices at UE, IRS and BS, $\mathbf{F}$, $\mathbf{P}$ and BS $\mathbf{W}$, \\
		the codebook matrices $\mathbf{A}_{\rm r}$ and $\mathbf{A}_{\rm t}$ \\
		\textbf{Initialization}: ${\Pi}_{r} = \varnothing$, ${\Pi}_{t} = \varnothing$, $\mathbf{B}_{\rm r} = \left( \mathbf{F}^{\rm T} \otimes \mathbf{W}^{\rm H} \right)\mathbf{A}_{\rm r}$, $\mathbf{B}_{\rm t} = \mathbf{A}_{\rm t} \mathbf{P}$\\
		1.~\textbf{Stage~1}: Estimate non-zero grid points in $\mathbf{A}_{\rm r}$ \\
		2.~~~~~$\mathbf{y}_{\rm sumr} =\sum_{i=1}^{M} \left(\mathbf{Y}\mathbf{B}_{\rm t}^{\rm H} \right) (:,i)$\\
		3.~~~~~$\mathbf{y} =\mathbf{y}_{\rm sumr}$, $\mathbf{B} = \mathbf{B}_{\rm r}$, $I\propto K_uK_bN_p^{\rm UI}N_p^{\rm IB}$\\
		4.~~~~~Use \textbf{Algorithm~2} to obtain the estimated grid point ${\Pi}_{r}$\\
		5.~\textbf{Stage~2}: Estimate non-zero grid points in $\mathbf{A}_{\rm t}$ \\
		6.~~~~~$\mathbf{y}_{\rm sumt} =\left(\sum_{i=1}^{KN_p^{\rm UI}N_p^{\rm IB}}\left(\mathbf{B}_{\rm r}^{\rm H}\mathbf{Y}\right)({\Pi}_{r}(i),:)\right)^{\rm T}$\\
		7.~~~~~$\mathbf{y} =\mathbf{y}_{\rm sumt}$, $\mathbf{B} = \mathbf{B}_{\rm t}$, $I \propto K_mN_p^{\rm UI}N_p^{\rm IB}$ \\
		8.~~~~~Use \textbf{Algorithm~2} to obtain the estimated grid point ${\Pi}_{t}$\\
		9. \textbf{Stage~3}: Recover the channel matrix \\
		10.~~~~~$\hat{\mathbf{A}}_{\rm r} = \mathbf{B}_{\rm r}(:, {\Pi}_{r})$, $\hat{\mathbf{A}}_{\rm t} = \mathbf{B}_{\rm t}(:, {\Pi}_{t})$\\
		11.~~~~~$\hat{\boldsymbol{\Lambda}}({\Pi}_{r},{\Pi}_{t}) = \hat{\mathbf{A}}_{\rm r}^\dagger \mathbf{Y} \left( \hat{\mathbf{A}}_{\rm t}^\dagger\right)^{\rm H} $\\
		\textbf{Output}: Estimated channel $\hat{\mathbf{H}} = {\mathbf{A}}_{\rm r} \hat{\boldsymbol{\Lambda}} ({\mathbf{A}}_{\rm t})^{\rm H}$ \\
		\bottomrule
	\end{tabular}
	\vspace{-5mm}
\end{table}

\subsubsection{SSE Algorithm}
The SSE algorithm separately estimates the positions of the non-zero grids on each side of the multiplied channel $\mathbf{H}^{\rm mul}$ in~\eqref{equ_Hmul_ongrid}. 
Specifically, since the non-zero grids on the left- and right-hand side codebook matrices $\mathbf{A}_{\rm r}$ and $\mathbf{A}_{\rm t}$ relate to the non-zero rows and columns of $\boldsymbol{\Lambda}$, respectively, we consider to separately estimate them. The procedures of the SSE algorithm are summarized in \textbf{Algorithm~1} and explained as follows.

At Stage~1, the non-zero grid points $\Pi_{\rm r}$ in $\mathbf{A}_{\rm r}$ is estimated. 
Specifically, by adding the columns of $\mathbf{Y}$ in Step~2, $\mathbf{y}_{\rm sumr}\in\mathbb{C}^{N_{sb}T_uT_b}$ can be expressed as $\mathbf{y}_{\rm sumr} = \left\vert \left( \mathbf{F}^{\rm T} \otimes \mathbf{W}^{\rm H} \right) \mathbf{A}_{\rm r} \mathbf{s}_{\rm sumr} + \mathbf{n}_{\rm sumr} \right\vert \in\mathbb{C}^{N_{sb} T_uT_b} $, where $ \mathbf{s}_{\rm sumr} = \left( \sum_{i=1}^{T_i}({\boldsymbol{\Lambda}} {\mathbf{A}}_{\rm t} \mathbf{P}\mathbf{P}^{\rm H} \mathbf{A}_{\rm t}^{\rm H})(:,i) \right) \in\mathbb{C}^{N_u N_b}$ denotes the equivalent transmit signal, and $\mathbf{n}_{\rm sumr}= \left( \sum_{i=1}^{T_i} \left( \mathbf{N } \mathbf{P}^{\rm H} \mathbf{A}_{\rm t}^{\rm H} \right) (:,i) \right) \in\mathbb{C}^{N_{sb}T_uT_b}$ refers to the equivalent noise.
Due to the sparsity of $\boldsymbol{\Lambda}$, $\mathbf{s}_{\rm sumr}$ is a sparse vector, the non-zero positions in $\mathbf{s}_{\rm sumr}$ relates to the non-zero rows of $\boldsymbol{\Lambda}$. Therefore, the positions of non-zero rows of $\boldsymbol{\Lambda}$ can be determined by estimating the non-zero positions of $\mathbf{s}_{\rm sumr}$, which is completed in Step~4.

Similarly, at Stage~2, the non-zero grid points $\Pi_{\rm t}$ in $\mathbf{A}_{\rm t}$ is estimated. 
Since the positions of the non-zero rows of $\boldsymbol{\Lambda}$ have been determined in the previous stage, using these rows of $\mathbf{Y}$ to compose $\mathbf{y}_{\rm sumt}$ is enough in determining the non-zero columns of $\boldsymbol{\Lambda}$, which is shown in Step~6. 
Moreover, $\Pi_{\rm t}$ is also determined by \textbf{Algorithm~2} in Step~8. 
Followed by that, at Stage~3 of \textbf{Algorithm~1}, the estimated ${\mathbf{A}}_{\rm r} $ and ${\mathbf{A}}_{\rm t} $ is first obtained in Step~10. 
The sparse on-grid channel matrix is then estimated in Step~11. 
Based on these estimated matrices, the channel matrix is finally recovered as illustrated in Step~11, which completes \textbf{Algorithm~1}.

To estimate the positions of non-zero grids with received signal $\mathbf{y}$ and measurement matrix $\mathbf{B}$, \textbf{Algorithm~2} first calculates the correlation between $\mathbf{B}$ and the residual vector $\mathbf{r}$ in Step~2. 
The most correlative column index is expressed as $n$, which is regarded as the newly founded grid index and added to the grid set $\Pi$. 
The estimated signal $\hat{\mathbf{s}}$ on the grids specified by $\Pi$ is calculated in Step~4 by using the LS algorithm. Then, the residual vector is updated in Step~5, by removing the effect of the non-zero grid points that have been estimated in the previous step. 
By repeating these procedures, $T$ indexes are selected as the estimated non-zero grid points.

\begin{table}[t]
	\centering
	\renewcommand
	\arraystretch{} 
	\begin{tabular}{l}
		\toprule
		\textbf{Algorithm 2:} Grid Position Estimation \\
		\midrule 
		\textbf{Input}: Received signal $\mathbf{y}$, measurement matrix $\mathbf{B}$, number of iterations $I$\\
		\textbf{Initialization}: $\Pi = \varnothing$, $\mathbf{r} = \mathbf{y}$, $\hat{\mathbf{s}} = \mathbf{0}_{{\text{size}}(\mathbf{B},2)}$\\
		1.~\textbf{for} $i = 1,\dots, I $\\
		2.~~~~~$n=\mathrm{argmax}~\Vert \mathbf{B}^{\rm H} \mathbf{r} \Vert_2^2$\\
		3.~~~~~${\Pi} = {\Pi} \cup n_r$\\
		4.~~~~~$\hat{\mathbf{s}}({\Pi}) = \mathbf{B}^{\dagger}(:,{\Pi})\mathbf{y} $\\
		5.~~~~~$\mathbf{r} = \mathbf{y} -\mathbf{B}\hat{\mathbf{s}} $\\
		6.~\textbf{end for}\\
		\textbf{Output}: Estimated grid position ${\Pi}$ \\
		\bottomrule
	\end{tabular}
	\vspace{-5mm}
\end{table}

\subsubsection{DSE Algorithm}

The computational complexity of the DSE algorithm majorly comes from the production in Step~2 of \textbf{Algorithm~2} in Step~4 and Step~8 of \textbf{Algorithm~1}, which are around $ \mathcal{O}\left(N_{sb}T_uT_b N_uN_b) \right)$ and $ \mathcal{O}\left( T_iM \right)$ in each iteration, respectively. 
These values become large with the increased number of antennas in the UM-MIMO and elements in the IRS. 
The DSE algorithm addresses this problem by exploiting the spatial correlation among subarrays. 
Specifically, in the HSPM channel~\eqref{equ_HSPM}, for the entire array on the left-hand side, the spatial angles from subarrays on the right-hand side are close.
Therefore, if we separately consider the codebooks between each subarray on the right-hand side and the entire array on the left-hand side, the positions of the non-zero grids would be close. 
Inspired by this, the DSE algorithm first calculates the positions of the non-zero grids for the codebook between the first subarray on the right-hand side and the entire array on the left-hand side, which are saved as the benchmark grids. 
For the remaining subarrays at the right-hand side, the grid searching space is shrunk by limiting the potential grids in the neighbor of the benchmark grids for reduced complexity. 

The grid shrinkage of the DSE algorithm operates at Stage~1 and Stage~2 of \textbf{Algorithm~1}, which are detailed in \textbf{Algorithm~3}. 
The input to the DSE algorithm is the summarized channel observation $\mathbf{y}_{\rm sum}$, the sensing matrix $\boldsymbol{\Phi}$, the codebook relating to the subarray at Tx, and the entire subarray at Rx $\mathbf{A}_{\rm sub}$, number of iterations $I$ and number of subarrays at right-hand side $K$.
At Stage~1 of \textbf{Algorithm~1}, these parameters are obtained as $\mathbf{y}_{\rm sum} = \mathbf{y}_{\rm sumr}$, $\boldsymbol{\Phi}= \mathbf{F}^{\rm T} \otimes \mathbf{W}^{\rm H}$, $\mathbf{A}_{\rm sub} = {\mathbf{U}}_{\rm u}^* \otimes \overline{\mathbf{A}}_{\rm rIRS-BS}$, $I \propto K_bN_p^{\rm UI}N_p^{\rm IB}$ and $K = K_u$, where $\mathbf{U}_{\rm u}$ denotes the spatial DFT matrix for the subarray at UE. 
At Stage~2 of \textbf{Algorithm~1}, these parameters are calculated as $\mathbf{y}_{\rm sum} = \mathbf{y}_{\rm sumt}$, $\boldsymbol{\Phi} = \mathbf{P}^{\rm T}$, $\mathbf{A}_{\rm sub} = {\mathbf{U}}_{\rm m}^* \otimes \overline{\mathbf{A}}_{\rm rUE-IRS}$, $I \propto K_mN_p^{\rm UI}N_p^{\rm IB}$ and $K = K_m$, where ${\mathbf{U}}_{\rm m}$ refers to the spatial DFT matrix for the subarray at IRS. 

\begin{table}[t]
	\centering
	\renewcommand
	\arraystretch{} 
	\begin{tabular}{l}
		\toprule
		\textbf{Algorithm 3:} DSE Algorithm for Grid Position Estimation\\
		\midrule 
		\textbf{Input}: Received signal $\mathbf{y}_{\rm sum}$, sensing matrix $\boldsymbol{\Phi}_{\rm}$, codebook matrix for the subarray $\mathbf{A}_{\rm sub}$
		\\number of iterations $I$, number of subarrays $K$\\
		\textbf{Initialization}: ${\Pi} = \varnothing$, ${\Pi}_{r,k} = \varnothing$, $N_a = \textrm{size}(\mathbf{A}_{\rm sub},1) $\\
		1.~\textbf{for} $k = 1:K$ \\
		2.~~~~~$\mathbf{Q} = \boldsymbol{\Phi}( :,(k - 1) * N_a+ 1 :k N_a)$\\
		3.~~~~~$\mathbf{B} = \mathbf{Q}\mathbf{A}_{\rm sub}$\\
		4.~~~~~\textbf{if} $k >1$ \\
		5.~~~~~~~~~$\mathbf{B} = \mathbf{Q}\mathbf{A}_{\rm sub}(:, \tilde{\Pi}) $ \\
		6.~~~~~\textbf{end~if}\\
		7.~~~~~Use \textbf{Algorithm~2} to obtain the estimated grid point ${\Pi}_{k}$\\
		8.~~~~~\textbf{if} $k=1$ \\
		9.~~~~~~~~Construct $\tilde{\Pi}$ by selecting the neighboring $q$ grids for each point in $\Pi_{1}$\\
		10.~~~~\textbf{end~if}\\
		11.~~~~Transform positions in ${\Pi}_{k}$ to positions in ${\Pi}$, ${\Pi} = {\Pi} \cup {\Pi}_{k}$\\
		12.~\textbf{end~for}\\
		\textbf{Output}: Estimated grid position ${\Pi}$\\
		\bottomrule
	\end{tabular}
	\vspace{-5mm}
\end{table}

For the $k^{\rm th}$ subarray on right-hand side, the DSE algorithm first obtains the sensing matrix $\mathbf{Q}$ and the corresponding measurement matrix $\mathbf{B}$, which are illustrated in Step~2 and Step~3 of \textbf{Algorithm~3}, respectively. 
For the first subarray, the non-zero grids relating to $\mathbf{A}_{\rm sub}$ are directly estimated and recorded in $\Pi_{1}$ as the benchmark grids. 
The neighboring $q$ elements for each grid in $\Pi_{1}$ are then selected as the potential searching grids for the remaining subarrays, which are saved as $\tilde{\Pi}$, as shown in Step~7 to 10 in \textbf{Algorithm~3}. 
For the remaining subarray pairs, only the grids in $\tilde{\Pi}$ will be searched, as illustrated in Step~4 to 6. 
Finally, in Step~11, the determined grid positions for subarrays are transformed to positions for the entire array and saved in $\Pi$. 

\subsubsection{Computational Complexity}
For the SSE algorithm, the total computational complexity of the SSE algorithm can be approximated as $ \mathcal{O}\left(I (N_{sb}T_uT_b N_uN_b  + T_iM)  \right)$. 
The computational complexity of the DSE algorithm also mainly comes from Step~4 and Step~8 of~\textbf{Algorithm~1}, the total computational complexity of the SSE algorithm can be approximated as $ \mathcal{O}\left(I \left(  \frac{N_{sb}T_uT_b N_uN_b}{K_u}  + \frac{T_iM}{K_m} \right)  \right)$.


\section{Performance Evaluation}
\label{sec_Performance_Evaluation}

In this section, we first numerically assess the system capacities by deploying different channel models for the THz integrated UM-MIMO and IRS systems. Then,
the performance of the proposed SSE and DSE CE algorithms is extensively evaluated. 

\subsection{Simulation Setup}
The simulation parameters and important notations used in this paper are summarized in TABLE~\ref{Tab_Simulation_Para}. 
We employ the system in Fig.~\ref{fig_system_model}, 
where the complex gain of the THz channel is generated based on the channel model in~\cite{ref_Multiray}. 
To evaluate the capacity, the IRS beamforming matrix $\overline{\mathbf{P}}$ in~\eqref{equ_received_signal_IRS} is randomly generated, while the phase of each element of $\overline{\mathbf{P}}$ follows a uniform distribution over $[0,2\pi]$.
In the CE process, we adopt the HSPM channel model in~\eqref{equ_HSPM} for both segmented channels $\mathbf{H}_{\rm UE-IRS}$ and $\mathbf{H}_{\rm IRS-BS}$. 
The spatial angles of both azimuth and elevation in the HSPM are randomly generated, following uniform distributions over $[0,\pi]$. 
The training process from~\eqref{equ_received_signal_biu} to~\eqref{equ_received_signal_IRS_training} are deployed. 
Specifically, the phase of each element of $\mathbf{W}_{b}$, $\mathbf{p}_{i}$ and $\tilde{\mathbf{F}}_{u}$ are randomly generated, following uniform distribution over $[0,2\pi]$.
The training time slot for BS, UE and IRS $T_b$, $T_u$ and $T_i$ satisfy $T_b \leq \frac{N_b}{K_b}$, $T_u \leq \frac{N_u}{K_u}$ and $T_i \leq M$ during our evaluation, to guarantee reduced training overhead.
The estimation accuracy is evaluated by the normalized-mean-square-error (NMSE), which is defined as 
\begin{equation}
	{\rm NMSE} = {\frac{ \mathbb{E} \left\{ \left\Vert \hat{\mathbf{H}} - \mathbf{H}^{\rm mul} \right\Vert_2^2 \right\} } { \mathbb{E} \left\{ { \left\Vert \mathbf{H}^{\rm mul} \right\Vert_2^2} \right\} }},
\end{equation}
where $\hat{\mathbf{H}}$ denotes the estimated channel. All the results are obtained by averaging 5000 trials of Monte Carlo simulations. 

\begin{table}[t]
	\centering
	\caption{Simulation parameters and notations.}
	\begin{tabular}{ccc}
		\toprule
		Notation& Meaning &Value in simulation\\
		\midrule
		$f$ & Carrier frequency & 0.3~THz \\
		$B$ & Bandwidth & 5~GHz\\
		$\lambda$ & 
		Carrier wavelength & \\
		$K_b, K_m, K_u$& Number of subarrays at the BS, IRS, and UE & Selected in 1,4\\
		$N_{ab},N_{am}$ & Number of antennas on a subarray at the BS and IRS & Selected in 64, 256\\
		$N_{au}$ & Number of antennas on a subarray at the UE &16 \\
		$N_b, M, N_u$ & Number of antennas at the BS, IRS and UE & \\
		$N_p^{\rm UI},N_p^{\rm IB}$ & Number of paths in $\mathbf{H}_{\rm UE-IRS}$ and $\mathbf{H}_{\rm IRS-BS}$  & $ N_p$\\
		$q$ & 
		Number of neighboring grids in the DSE algorithm & 5 \\
		$T_{u}, T_b, T_i$ & Training time slots of UE, BS and IRS & \\
		$N_r, N_t$& Number of antennas at Rx (IRS or BS) and Tx (UE or IRS) & \\
		$K_r, K_t$& Number of subarrays at Rx (IRS or BS) and Tx (UE or IRS)& \\
		
		$\mathbf{A}_{\rm r}, {\mathbf{A}}_{\rm t}$ & Codebook matrices at left and right side of $\mathbf{H}^{\rm mul} $, respectively & \\
		$\tilde{\mathbf{F}}_{u}, \mathbf{W}_{b},\mathbf{p}_{i}$ & 
		Training beamforming, combining and IRS reflection matrices & \\
		${\mathbf{F}}, \mathbf{W},\mathbf{P}$ & 
		Combined training beamforming, combining, IRS reflection matrices & \\
		$\overline{\mathbf{F}},\overline{\mathbf{W}},\overline{\mathbf{P}}$& UE beamforming, IRS beamforming and BS combining matrices & \\	
		$\mathbf{H}_{\rm UE-IRS}, \mathbf{H}_{\rm IRS-BS}$ & Segmented channels form UE to IRS and IRS to BS, respectively\\
		$\mathbf{H}_{\rm P}, \mathbf{H}_{\rm S}, \mathbf{H}_{\rm HSPM}$ & PWM, SWM and HSPM channel matrices &\\
		$\mathbf{H}^{\rm cas}_{\rm PWM}, \mathbf{H}^{\rm cas}_{\rm SWM}, \mathbf{H}^{\rm cas}_{\rm HSPM}$ & Cascaded channels based on the PWM, SWM and HSPM &\\
		$\mathbf{H}^{\rm mul} $ & The multiplied channel matrix to be estimated in~\eqref{equ_received_signal_IRS_training}& \\
		$\mathbf{Y}$ & Observation matrix used for CE after training in~\eqref{equ_received_signal_IRS_training} & \\	
		${\boldsymbol{\Lambda}}$ & Sparse on-grid channel of $\mathbf{H}^{\rm mul} $ based on $\mathbf{A}_{\rm r}$ and ${\mathbf{A}}_{\rm t}$\\
		\bottomrule
	\end{tabular}
	\vspace{-5mm}
	\label{Tab_Simulation_Para}
\end{table}

\subsection{System Capacity based on Different Channel Models}

We begin by evaluating the system capacities by using PWM, SWM, and HSPM for the segmented channels in different communication distances and subarray spacing. 
To facilitate evaluation, we consider that both cascaded channels $\mathbf{H}_{\rm UE-IRS}$ and $\mathbf{H}_{\rm IRS-BS}$ have only one LoS path, i.e. $N_p=1$, which is simplified yet practical for the THz communication systems due to the LoS domination property~\cite{ref_THz_LoS_MIMO}. In this case, the $\mathbf{H}^{\rm{cas}}_{\rm PWM}$ has rank 1 with no spatial multiplexing capability. 
Moreover, the transmit power at the BS is fixed as 20 dBm.

\begin{figure}[t]
	\setlength{\belowcaptionskip}{0pt}
	\centering
	\subfigure[$N_u = 64, M= N_b = 256, K_u = K_m = K_b = 4$.]{
		\includegraphics[width=2.8in]{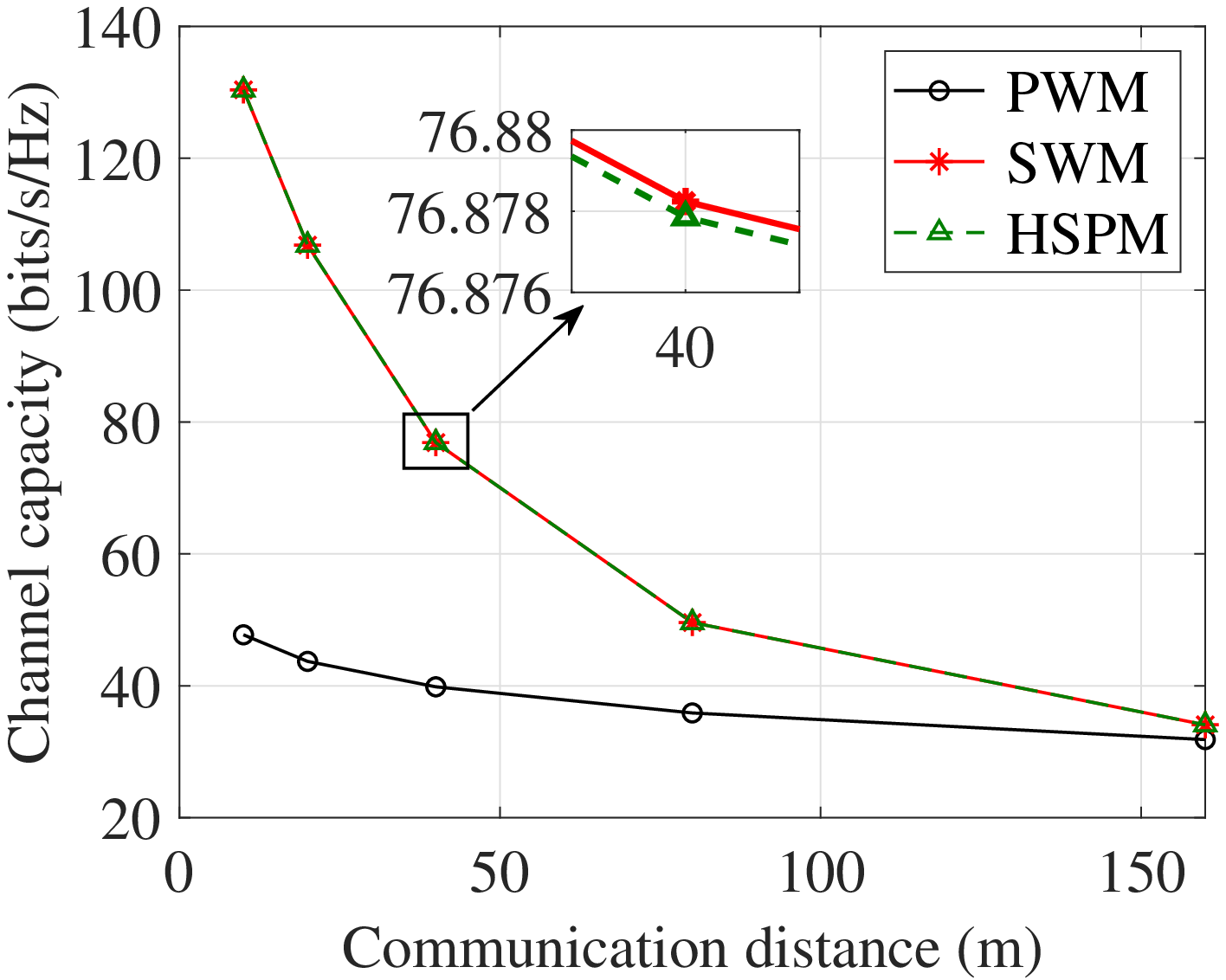} }
 \subfigure[$N_u = 64, M =N_b = 1024, K_u = K_m = K_b = 4$.]{
		\includegraphics[width=2.8in]{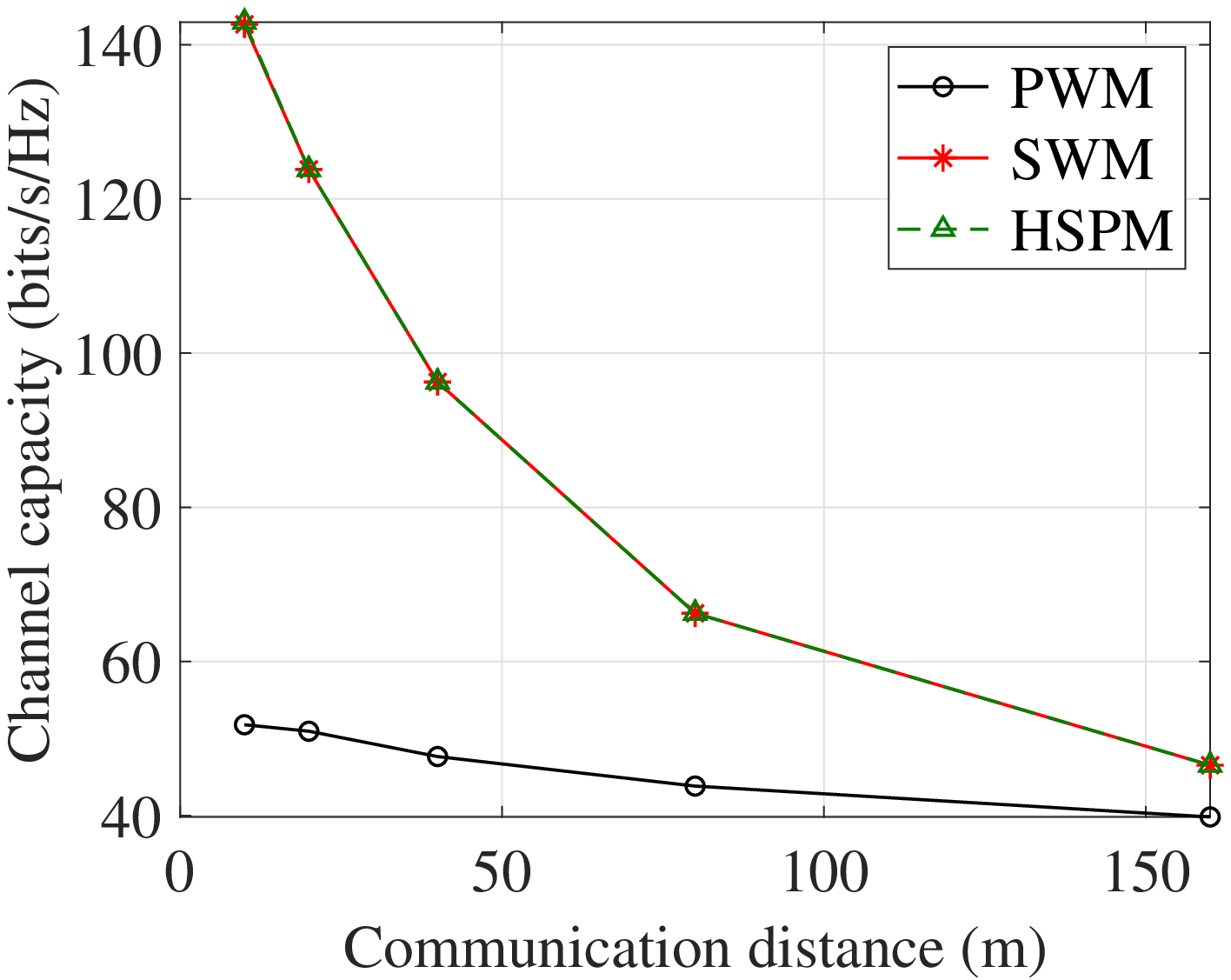} }
		\caption{Channel capacity with various communication distance, the subarray spacing is fixed as $64\lambda$.}
		\label{fig_RIS_Cap_dist}
		\vspace{ -5mm}
\end{figure}

The capacity results over different communication distances from BS to IRS and IRS to UE are illustrated in Fig.~\ref{fig_RIS_Cap_dist}.
It is observed that the capacity of $\mathbf{H}^{\rm cas}_{\rm HSPM}$ is very close to $\mathbf{H}^{\rm cas}_{\rm SWM}$, which is much higher than that based on $\mathbf{H}^{\rm cas}_{\rm PWM}$. In particular, as shown in Fig.~\ref{fig_RIS_Cap_dist}(a), when the communication distance is 40 m, the capacity of $\mathbf{H}^{\rm cas}_{\rm HSPM}$ is only $5\times 10^{-4}$ bits/s/Hz lower than that of $\mathbf{H}^{\rm cas}_{\rm SWM}$.
The capacities of $\mathbf{H}^{\rm cas}_{\rm HSPM}$ and $\mathbf{H}^{\rm cas}_{\rm SWM}$ are $37.0$ bits/s/Hz higher than the capacity of $\mathbf{H}^{\rm cas}_{\rm PWM}$.
This is explained that in near-field transmission, the PWM loses its effectiveness to characterize the channel. 
By contrast, the capacities based on PWM, HSPM and SWM converge when the communication distance far exceeds the Rayleigh distance, which equals to 46.3~m in this case.
Particularly, when the communication distance is 160~m that is over three times of the Rayleigh distance, the capacities based on different channel models finally approach to be close, where the difference between the PWM and the other channels reduces to 2.2 bits/s/Hz.
Therefore, the Rayleigh distance overestimates the accuracy of the PWM approximation from the SWM. Equivalently, the misuse of PWM could cause severe deterioration of capacity even when the communication distance is equal to or larger than the Rayleigh distance, i.e., the so called far-field region. 
As a take-away lesson from our analysis, the HSPM is effective and generally applicable  when the communication distance is smaller, comparable or even larger than the Rayleigh distance, i.e., ranging from near-field to far-field.

\begin{figure}[t]
	\setlength{\belowcaptionskip}{0pt}
	\centering
	\subfigure[$N_u = 64, M= N_b = 256, K_u = K_m = K_b = 4$]{
		\includegraphics[width=2.8in]{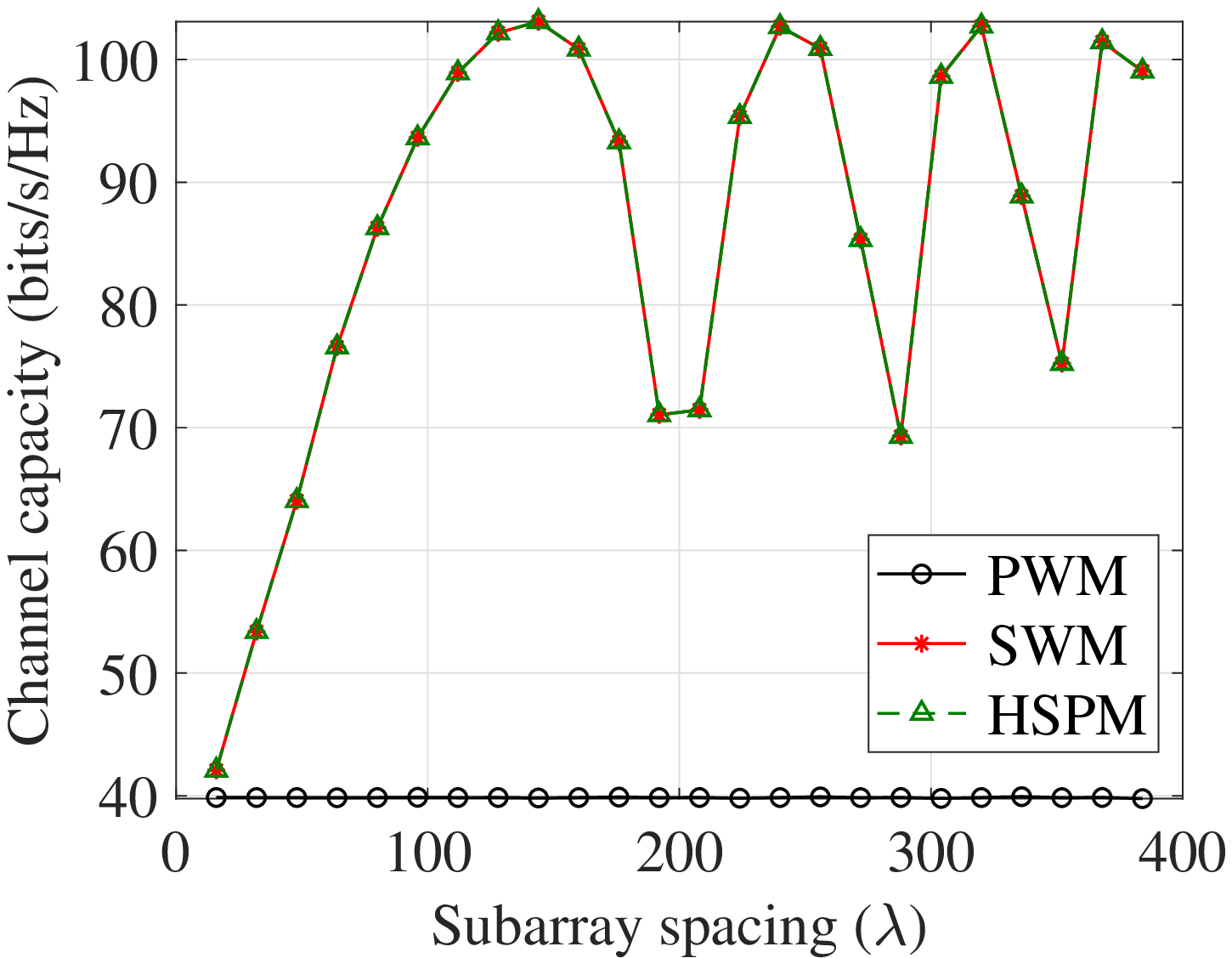} }
 \subfigure[$N_u = 64, M =N_b = 1024, K_u = K_m =K_b = 4$.]{
 
		\includegraphics[width=2.8in]{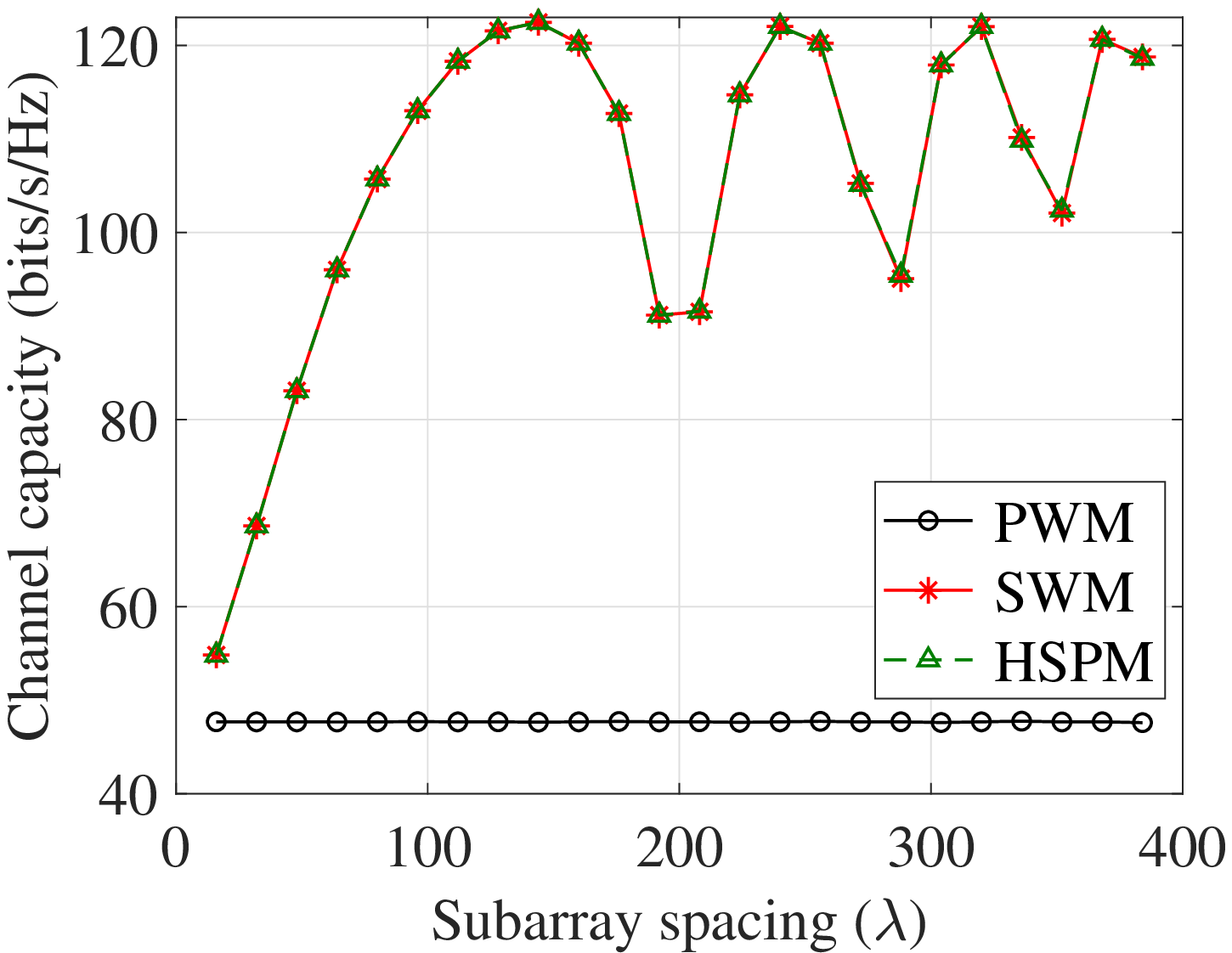} }
		\caption{Channel capacity with various subarray spacing, the communication distance is fixed as $40$ m. }
		\label{fig_RIS_Cap_suba_dist}
		\vspace{ -5mm}
	\end{figure} 

As illustrated in Fig.~\ref{fig_RIS_Cap_suba_dist}, the effect of subarray spacing on channel capacity is evaluated with varying numbers of elements in the UM-MIMO and IRS.
The trends of the curves in Fig.~\ref{fig_RIS_Cap_suba_dist}(a) and Fig.~\ref{fig_RIS_Cap_suba_dist}(b) are identical, due to the similar array size, which is mainly dependent on the subarray spacing. Specifically, in the considered system, the channel capacity is majorly influenced by the condition number, i.e., the difference between the minimax eigenvalues for the channel.
As studied in~\cite{ref_LoS_MIMO_rank}, with fixed communication distance, the eigenvalue is a function of the array size. 
Moreover, when the subarray spacing is smaller than a threshold, e.g., $144\lambda$ in both figures, the channel capacity rises monotonically with larger subarray spacing. In particular, as illustrated in Fig.~\ref{fig_RIS_Cap_suba_dist}(a), the capacity increases from 42.0~bits/s/Hz to 103.1~bits/s/Hz for the HSPM and SWM, as the subarray spacing increases from $16\lambda$ to $144\lambda$.
This is explained that the enlarged subarray spacing expands the near-field region and provides a better condition number to the channel, which contributes to the spatial multiplexing gain~\cite{ref_HSPM}. By contrast, the capacity based on the PWM remains around 39.9~bits/s/Hz. 
In addition, as the subarray distance further increases beyond $144\lambda$, the capacity begins fluctuating, due to the variation of the eigenvalues
of the channel matrix~\cite{ref_LoS_MIMO_rank}.
In this study, we consider the reasonable widely-spaced subarrays, e.g., the subarray spacing is smaller than $144\lambda = 0.144$~m. Therefore, the spatial multiplexing of the THz integrated UM-MIMO and IRS systems can be improved based on the widely-spaced architecture design. 

\subsection{Performance of SSE and DSE Channel Estimation}



\begin{figure}[t]
	\setlength{\belowcaptionskip}{0pt}
	\centering
	\subfigure[WSMS system using HSPM. $N_u = 16, M = N_b = 1024, K_u = 1, K_m = K_b = 4, T_u = 12, T_i = 768, T_b = 192$]{
	\includegraphics[width=0.31\textwidth]{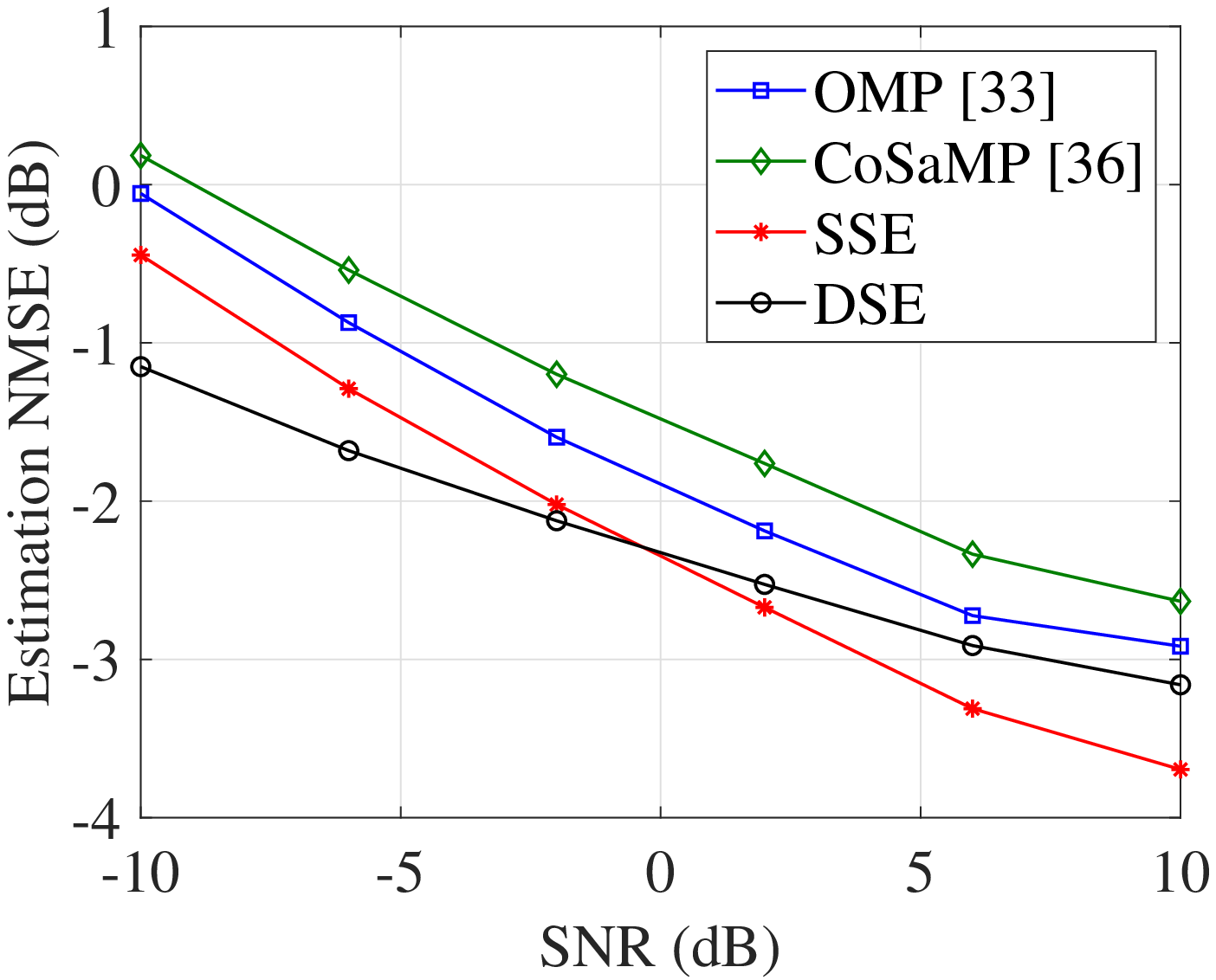} }
	\subfigure[WSMS system using SWM, $N_u = 16, M = N_b = 1024, K_u=1, K_m=K_b = 4, T_u = 12, T_i=192, T_b = 192$]{
	\includegraphics[width = 0.31\textwidth]{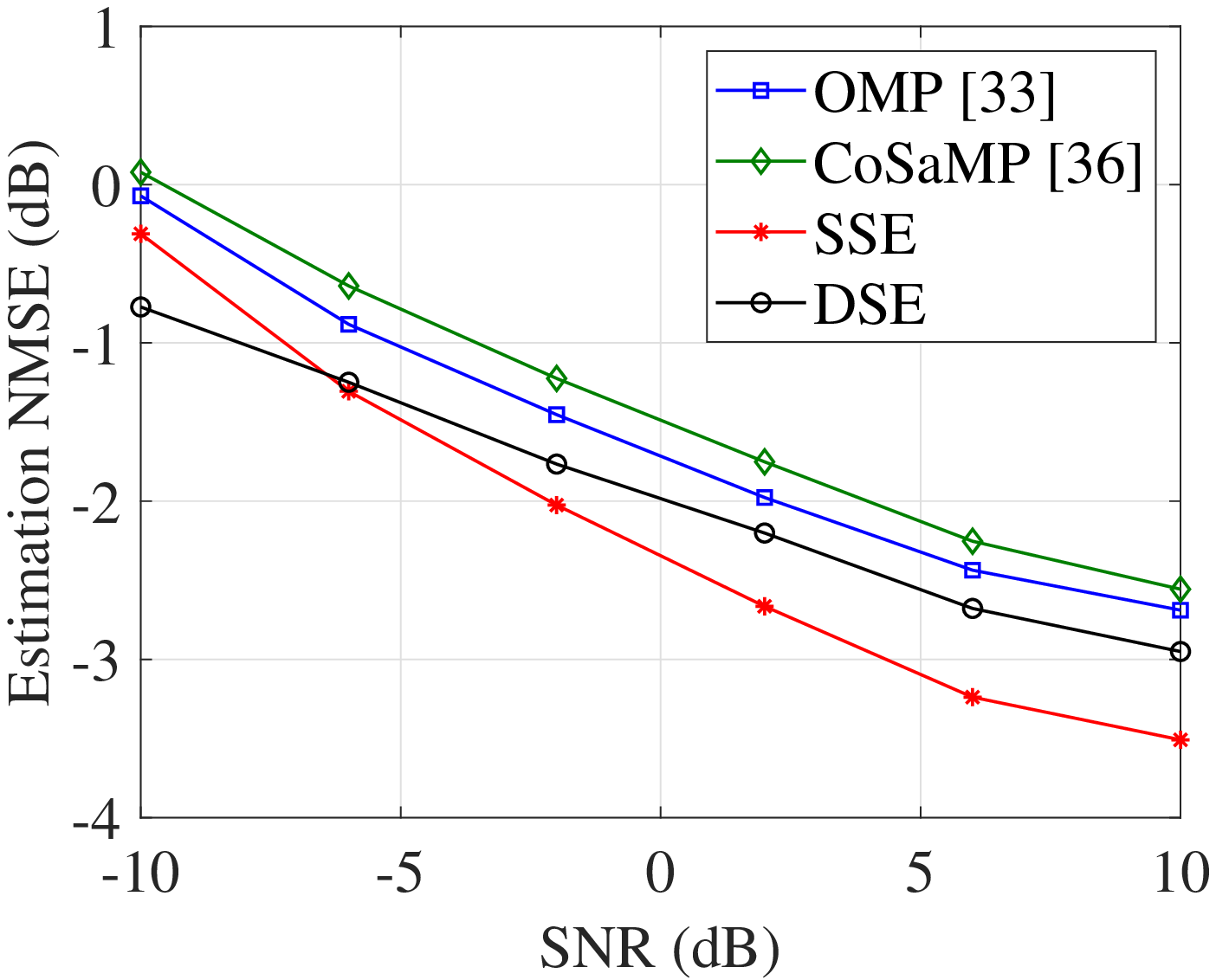} }
 	\subfigure[Compact array systems using PWM, $N_u = 16, M = N_b = 256, K_u=K_m=K_b = 1, T_u = 12, T_i=192, T_b = 192$]{
	\includegraphics[width = 0.31\textwidth]{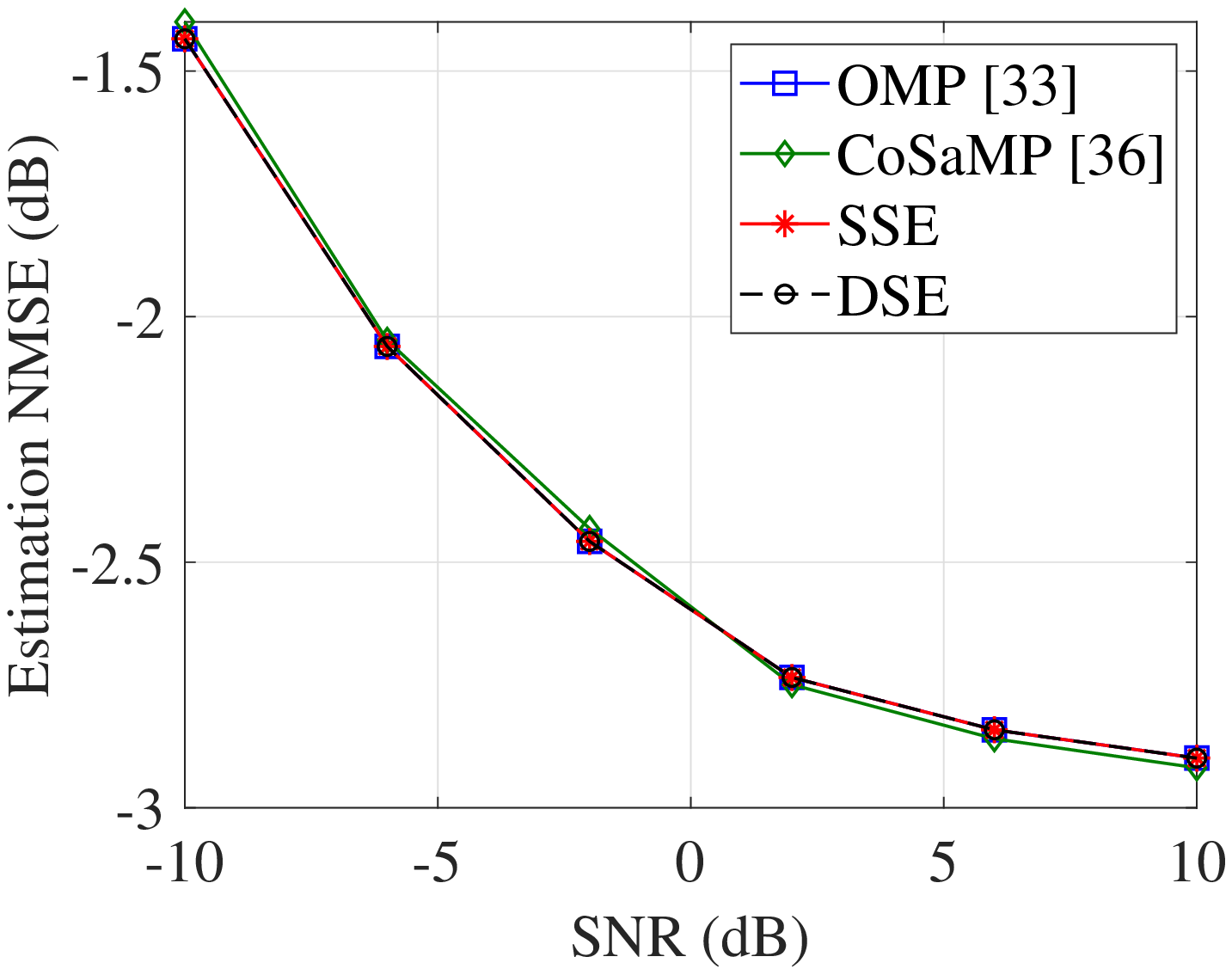} }
	\caption{NMSE comparison of different CE algorithms in different systems using different channel models. }
	\vspace{-5mm}
	\label{result_CE_channels}
\end{figure}

To demonstrate the effectiveness of the proposed subarray-based codebook, we first compare the NMSE performance of the proposed SSE and DSE algorithms with two classical on-grid CS-based algorithms in different systems by deploying different channel models, including the OMP method as in~\cite{ref_OMP_IRS} and the CoSaMP~\cite{ref_CoSaMP}, both of which deploy the traditional DFT codebook. 
In addition, we fix the number of paths as $N_p=2$ for each channel segment. 
As illustrated in Fig.~\ref{result_CE_channels}(a), the estimation NMSE against the SNR under the HSPM channel is evaluated. 
The proposed SSE and DSE methods based on the proposed codebook perform better than the traditional methods based on the DFT codebook.
This observation validates the accuracy and effectiveness of the proposed subarray-based codebook in the considered system. 
Moreover, at higher SNR values, i.e., SNR$>$0~dB, the SSE algorithm performs the best and obtains the highest estimation accuracy. Specifically, at SNR = 6~dB, the estimation NMSE of the SSE is 1~dB, 0.6~dB and 0.4~dB lower than the OMP, CoSaMP and DSE counterparts, respectively.

By contrast, at low SNR values, we can observe that the performance of the low-complexity DSE algorithm exceeds that of the SSE algorithm.
For instance, the estimation NMSE of the DSE is around 0.8~dB lower than that of the SSE at -10~dB SNR.
This gap decreases with the increment of SNR. The NMSE of SSE becomes lower than that of DSE as the SNR exceeds 0~dB. 
This is explained that, the potential grids error in the DSE algorithm can be avoided by the determination of potential searching grids based on the benchmark grids, especially in noisy conditions. 
However, since the best grids for the entire array cannot be completely mapped to the first subarray, the performance of the DSE becomes worse than the SSE as the SNR increases. 
To this end, we can state that the DSE algorithm is more attractive in the low SNR region, i.e., SNR$<$0~dB. 
Furthermore, by considering the same system configuration as in Fig.~\ref{result_CE_channels}(a), the estimation NMSE of different algorithms by deploying the ground-truth SWM is evaluated in Fig.~\ref{result_CE_channels}(b). 
The result is consistent with that in Fig.~\ref{result_CE_channels}(a), which further reinforces the effectiveness of the proposed HSPM. Specifically, the estimation accuracy of the SSE outperforms the other algorithms when at higher SNR larger than -5~dB, while the DSE algorithm achieves the lowest NMSE among the evaluated algorithms when SNR$<$-5~dB. 

To study the performance of the proposed SSE and DSE algorithms even in the traditional compact array systems without enlarging the subarray spacing, we evaluate their performances in Fig.~\ref{result_CE_channels}(c) in contrast to the OMP and CoSaMP algorithms. 
We observe that the estimation NMSE of the OMP, SSE, and DSE algorithms are close. 
This is explained that in the traditional compact array systems, the number of subarrays at the BS, UE, and IRS is equal to 1.
Therefore, the subarray-based codebook degenerates into the DFT codebook, and the operations in the DSE and SSE algorithms become the same.  
This result further demonstrates the effectiveness of the subarray-based codebook in the WSMS systems. 
Instead of being restricted by the performance of the sparse recovery algorithms, the performances of the OMP and CoSaMP methods are limited by the accuracy of the DFT codebook.
Although very close, the NMSE of the CoSaMP slightly outperforms the remaining algorithms as the SNR exceeds 0~dB, with 0.02~dB higher NMSE at SNR=10~dB. This is owing to the benefits of the grid selection mechanism in the CoSaMP~\cite{ref_CoSaMP}.

\begin{figure}[t]
	\setlength{\belowcaptionskip}{0pt}
	\centering
	\subfigure[NMSE against $T_i$, $N_u = 16, M = N_b = 1024, K_u = 1, K_m = K_b = 4, T_u = 16, T_b = 192$]{
		\includegraphics[width=2.8 in]{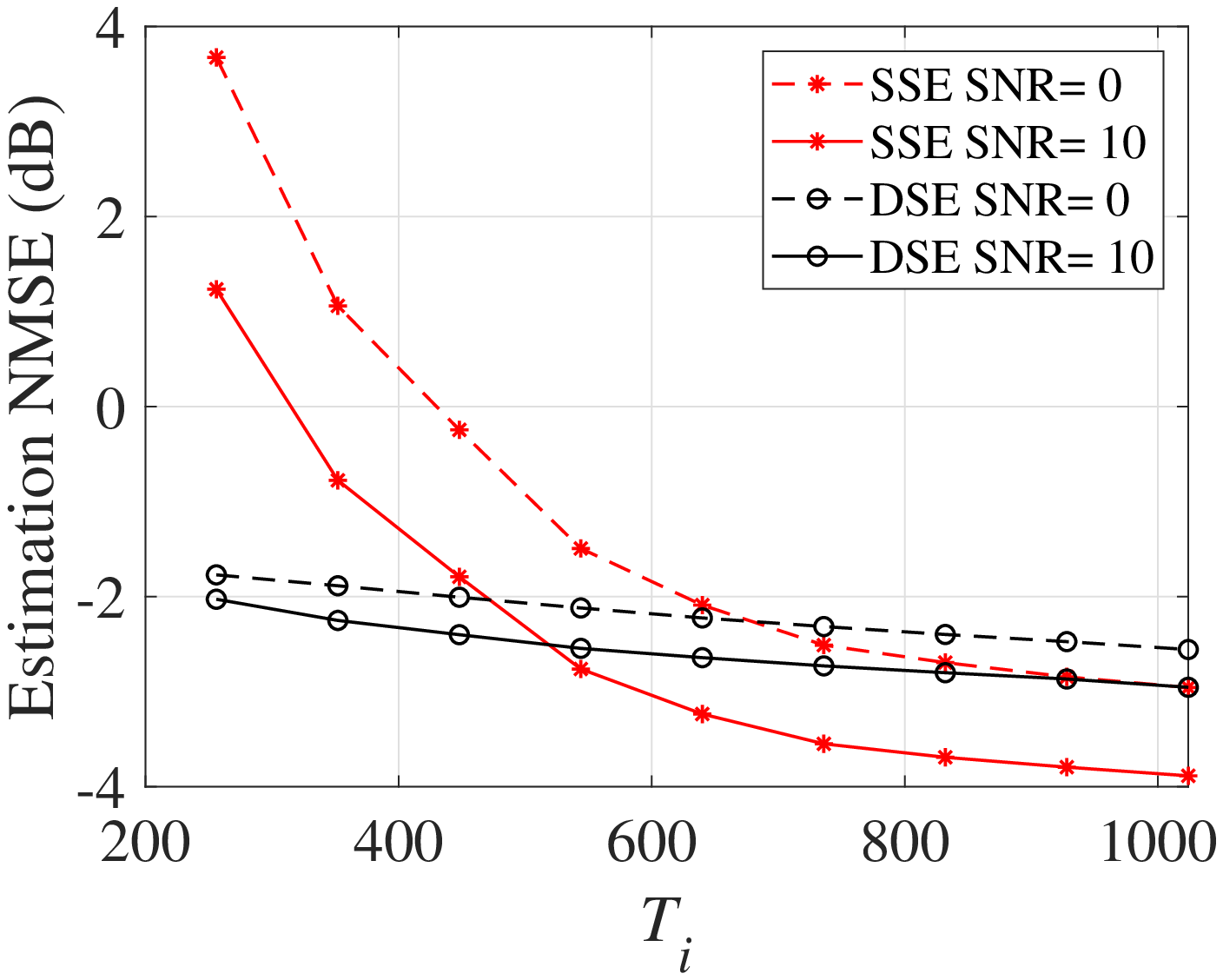} }
	\subfigure[NMSE against $T_b$, $N_u = 16, M = N_b = 1024, K_u =1, K_b = K_m = 4, T_u = 16,T_b = 192$.]{\includegraphics[width=2.8 in]{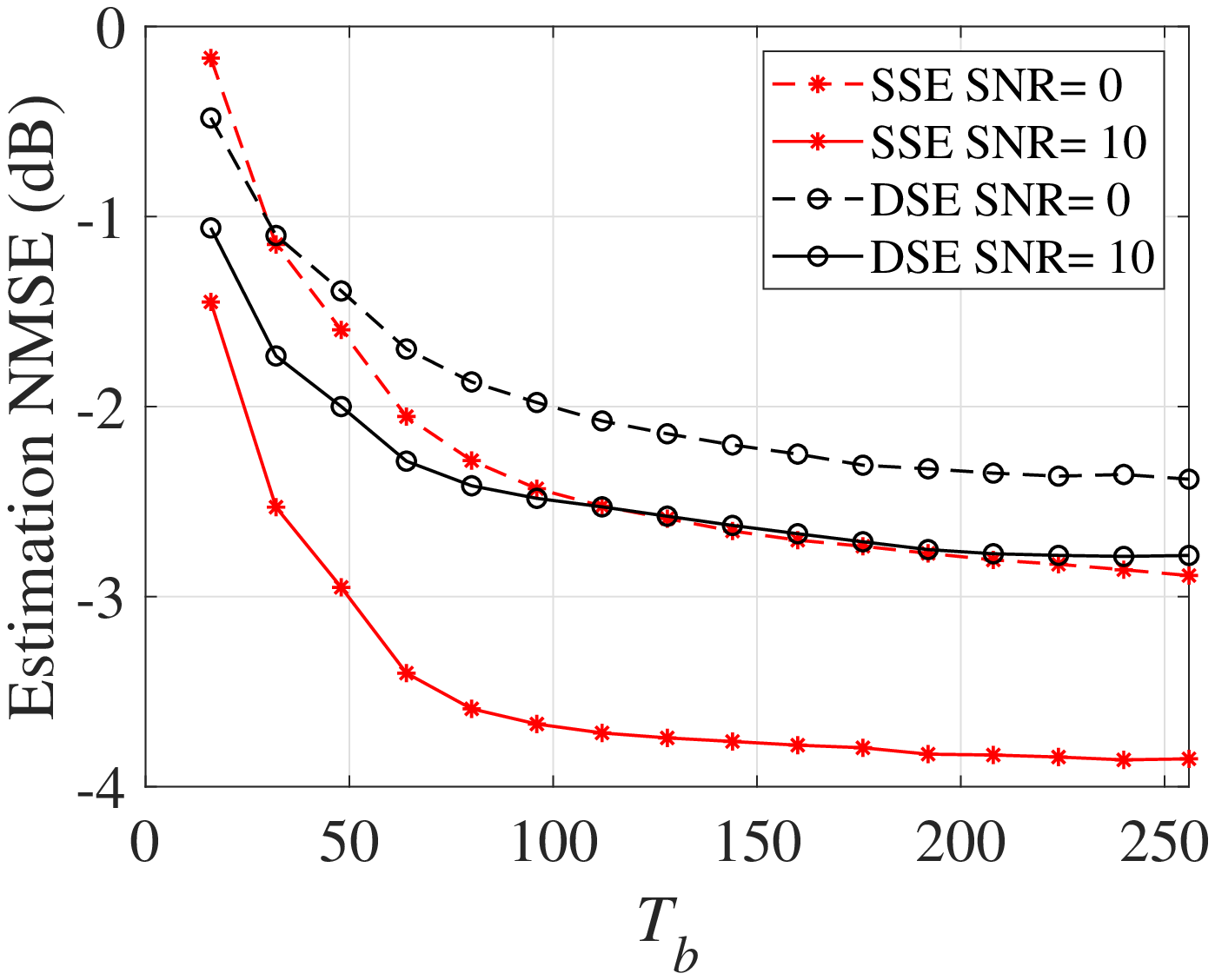} } \caption{Estimation NMSE against the number of training slots. } \vspace{-5mm} \label{result_CE_cv_pilot}
\end{figure} 

In Fig.~\ref{result_CE_cv_pilot}, the NMSE performances of the DSE and SSE schemes versus the length of training slot in different SNRs are analyzed. The NMSE decreases with the increased length of training slots allocated to both the IRS and BS. 
In particular, as illustrated in Fig.~\ref{result_CE_cv_pilot}(a), pertaining the SSE algorithm, when SNR is 10 dB, the NMSE drops sharply first, in which the decrement is around 4.5~dB as $T_i$ increases from 256 to 640. 
However, as $T_i$ further increases from 640 to 1024, the decline tends smooth, i.e., NMSE decreases by 0.65~dB. 
This is due to the fact that the NMSE results are determined by the dimensions of the channel and channel observation. The latter one is enlarged with the increment of the pilot length. 
Thus, the channel is more accurately estimated with a longer training slots, especially when the dimensions of the observation and the channel are comparable.
In addition, as shown in Fig.~\ref{result_CE_cv_pilot}(a) and Fig.~\ref{result_CE_cv_pilot}(b), to obtain a good CE performance, time slots of length 600 and 100 are enough for the IRS and BS training in the considered configuration, respectively, after which the NMSE degradation is very limited by adding the length of the training slots. 
These values take around only half of the required training time slots for the traditional least-square (LS) and minimum-mean-square (MMSE) CE methods~\cite{ref_trice}, in which $1024$ and $256$ slots are required for the IRS and BS training, respectively.
These numbers are obtained by $M$ and $\frac{N_b}{K_b}$, respectively. 
Therefore, the proposed CE framework can estimate the channel with reduced training overhead. 
\section{Conclusion}
\label{sec_Conclusion}
 
As a promising technology for THz communications, the integrated UM-MIMO and IRS systems can effectively solve the LoS blockage problem in complex occlusion environments.
Three challenges arise. 
First, the huge dimensional antenna array in UM-MIMO and IRS in contrast with the sub-millimeter wavelength enlarges the near-field region of propagation. 
Second, the spatial multiplexing and capacity are strongly limited by the sparsity of the THz channel.
Third, the adoption of hybrid beamforming systems in UM-MIMO results in limited RF-chains, with the lack of signal processing capability of the IRS, CE has to recover the high-dimensional channel from severely compressed observations. 

In this work, we have proposed the HSPM to accurately model the cascaded channel of the THz integrated UM-MIMO and IRS system. 
We have analyzed the spatial multiplexing under near-field and far-field cases and proved that the spatial multiplexing of the cascaded channel is limited by the segmented channel with a lower rank.
Additionally, we have developed a subarray-based on-grid codebook and the SSE and DSE algorithms with low complexity to address the CE problem. 
Extensive simulations are conducted, and results demonstrate the accuracy of the HSPM channel. The capacity based on HSPM is only $5\times10^{-4}$ bits/s/Hz lower than that based on the ground-truth SWM with an array size of $256$.
Moreover, the spatial multiplexing gain is improved based on the widely-spaced architecture design. 
Based on the proposed codebook, the SSE and DSE algorithms achieve better estimation accuracy than traditional algorithms. 
While the SSE possesses the highest accuracy at higher SNR over 0~dB, the DSE is more attractive at low SNR, whose estimation NMSE is 0.8~dB lower than the SSE when SNR=-10~dB. 

	\bibliographystyle{IEEEtran}
	\bibliography{reference} 
\end{document}